# Thermodynamic analysis for Non-linear system (Van-der-Waals EOS) with viscous cosmology

## Shouvik Sadhukhan[1], Alokananda Kar[2], Surajit Chattopadhay[3]

1. Department of Physics; Indian Institute of Technology ; Kharagpur-721302 ; West Bengal ; India
   Email:shouvikphysics1996@gmail.com
2. Department of Physics; University of Calcutta; 92 APC ROAD Kolkata-700009 ; West Bengal ; India
   Email:alokanandakar@gmail.com
3. Department of Mathematics, Amity University Major Arterial Road, Action Area II, Rajarhat New Town, Kolkata 700135, India
   *Corresponding author Email:surajitchatto@outlook.com

**Abstract**

The following paper has been motivated from the recent works of Kremer, G.M [1, 4], Vardiashvili, G [2], Jantsch, R.C [3], Capozziello, S [5] on Van-Der-Waals EOS cosmology. The main aim of this paper is to analyze the thermodynamics of a Non-linear system which in this case is Van-Der-Waals fluid EOS [6]. We have investigated the Van-Der-Waals fluid system with the generalized EOS as $p = w(\rho,t)\rho + f(\rho) - 3\eta(H,t)H$ [8]. The third term signifies viscosity which has been considered as an external parameter that only modifies pressure but not the density of the liquid. The $w(\rho,t)$ and $f(\rho)$ are the two functions of energy density and time that are different for the 3 types of Vander Waal models namely one parameter model, two parameters model and three parameters model [7, 9]. The value of EOS parameter ($w_{EOS}$) [6, 10] will show different values for different models. We have studied the changes in the parameters for different cosmic phases [4, 5, 6]. We have also studied the thermodynamics and the stability conditions for the three models in viscous condition [10, 12, 13, 14]. We have discussed the importance of viscosity [11] in explaining accelerating universe with negative pressure [12]. Finally, we have resolved the finite time future singularity problems [15-20] and discussed the thermodynamics energy conditions [21-31] with those models.

## 1. Introduction

The standard model in cosmology predicts several phases during the entire life-time of the universe. FLRW as well as other models present some interior cosmic phases that are necessary to resolve the cosmological problems. The cosmic inflation is one of them. The idea of cosmological inflation has been brought in physics to resolve the cosmological problems (viz. Vacuum energy density problem [50-52], magnetic monopole problem, cosmic Horizon problem and flatness problem). This theory of cosmological inflation has been established with both dark energy [43, 45] as well as with geometry. It was assumed that inflation is basically a phase transition between the primordial Planck scaled cosmology and present universe late time acceleration phase. Critical analysis of this phenomenon provides us with the knowledge of the decelerated phase of universe which is followed by inflation and reheating and further another phase transition to produce late time acceleration. The phase after reheating is followed by radiation dominated pressure. Other accelerated phases have been assumed to be controlled by negative pressure [17, 23, 24, 54, 55, 58] which is quite impossible with Einstein action and thus the alternative gravity [23-31] has been introduced. From Raychaudhuri equation we understand that strong energy condition is violated for accelerated expansion of universe with attractive gravity or positive pressure. Therefore, several steps have been taken in physics. Non-linearity and Inhomogeneity in cosmic fluid are some of them. [8, 32]

Cosmic fluid viscosity is another fundamental problem in cosmology. Local dissipation is the main cause to assume large scale global fluid dissipation. The negative pressure problem in universe evolution can be resolved with the introduction of viscosity. In some literature it is also assumed that the negativity in cosmic fluid pressure can be brought with the introduction of viscosity before cosmic inflation. Thus, positive viscosity becomes an important factor for inflation. Negative viscosity [11] is another idea used to introduce excess energy effect in cosmic fluid. Single fluid negative viscosity [11] violates the second law of thermodynamics and thus it brings the necessity of introduction of multi-fluid system in cosmology. The positive viscosity in cosmic fluid is excessively available in literature but negative viscosity is quite rare. The inhomogeneity in fluid [33-41] is mainly brought in physics with viscosity that modifies the pressure and contributes itself in the overall evolution.

Dark energy is the most successful alternative gravity model that has discussed the negative pressure and cosmic inflation problems. There are three different types of dark energy models (viz. fluid model, scalar field, holographic model). These models are also successful to discuss the vacuum energy problem, late time acceleration as well as the finite time future singularity problems [15-20, 32]. Dark energy also substituted with modified gravity with higher order correction of Einstein action. Phantom [34], Quintom, Quintessence [6, 43, 45, 46, 47], K-Essence, tachyon [42, 45] and DBI-Essence models are included into scalar field models. Holographic models [22, 42, 54, 57, 59] include the IR-UV cutoff of cosmological constants. Example of fluid Dark energy model is Chaplygin gas [12, 13]. Here we have used the non-linear fluid system to substitute the results of Dark energy and modified gravity [28].

Non-linearity in the fluid of cosmology is nothing but the modification of equation of state with higher order correction [1-5]. This correction might provide the solution of the negative pressure problem, late time acceleration and cosmic acceleration problems [6, 7]. Several literatures are available to analyze the non-linear models. The non-linear models (EOS) can also produce inflation and late time acceleration with the help of viscosity [8, 10]. The Van-Der-Waals gas model is one of the most generalized non-linear models. Van-Der-Waals gas equations can produce the equation of states according the following derivations [9].

The Van-Der-Waals model equation of state belongs to the Van-Der-Waals gas law in general thermodynamics. From that gas law we may write as follows.

$$\left(p + \frac{a}{V^2}\right)(V - b) = C$$

Here $C =$ Constant.

Now from above equation we get,

$$\frac{1}{\rho}(p + a\rho^2)(1 - b\rho) = C$$

Here we considered $\frac{1}{V} \propto \rho$. So, we get again,

$$p = \frac{C\rho}{(1 - b\rho)} - a\rho^2$$

Here $C, a$ and $b$ have different values for three models which will be subsequently discussed in the next section onwards. These values can be derived in terms of critical density, critical pressure as well as critical volume. In several literatures the coefficient of energy density of the above EOS is considered to be a function of energy density and the coefficient can be substituted with $w(t, \rho) = \frac{C}{1-b\rho}$. The details will be discussed in next section.

This paper is aimed to discuss the generalized form of the non-linear models with introduction of equation of state discussed in the above paragraph. Here we have discussed the basics of the three different types of EOS (three different form of the EOS derived from the above paragraph) with different parameters (viz. $w, \gamma, \rho_c, \alpha$ and $\beta$) [2]. In general, cosmology is controlled by the phase conditions provided by the EOS parameter but here our derivation has proved that the cosmic phases can be defined with other parameters also. These parameters together provide the necessary conditions of inflation and late time accelerations [5,6]. The fluid thermodynamics of cosmology has been discussed for these non-linear models [12-14].

In view of the literature survey presented above, we have attempted a thermodynamic analysis for Van-der-Waals EOS considering of viscous cosmology. Rest of the paper is organized as follows. In section 2 and 3 we discussed the thermodynamics of the Van-Der-Waals models with viscosity. In section 4 and 5 we have discussed the basics of thermodynamics energy conditions as well as the resolution of finite time future singularity problems for the models. In section 6 we have discussed the scalar field theory and discussed the inflation conditions. In section 7 and 8 we have discussed thermodynamics stability for viscous and non-viscous conditions respectively.

## 2. Overview of Van-der-Waals model of cosmic fluid

As already mentioned we will be using the van-der-Waals fluid with the form of EOS as discussed in paper of Brevik, I [8].

$$p_m = w(\rho, t)\rho + f(\rho) - 3\eta(H, t)H$$

(1)

Where $w(\rho, t)$ and $f(\rho)$ [9] are two functions of $\rho$, $P_m$ is the modified pressure due to the viscosity which reduces the actual pressure of the fluid. The EOS parameter $w(t, \rho)$ is assumed according to the EOS discussion given in the introduction section. The function $f(\rho)$ is taken instead of $\rho^2$ for generalization purpose (although we have considered $f(\rho) = constant \times \rho^2$ in later discussion). The third term of the equation of state equation 1 has been taken to establish the viscous case where the coefficient of viscosity is a function of time and Hubble parameter. This term actually converts it inhomogeneous where the second density dependent function makes it non-linear.

We have taken the form $\eta = \eta(t)(3H)^n$ and $\eta(t) = \tau$ as discussed in the work of [8].

### 2.1. Signification of these three different models

We have established a generalized non-linear equation of state from Van-Der-Waals fluid formalism that later transformed into three different models with different number of parameters [2]. The main significance behind these parameter is that they act like the EOS parameter [2]. In non-linear fluid models the EOS parameter provides smooth curves to discuss the cosmological phase transitions [6, 32, 51]. It is efficient to discuss the conditions for quintom phases [53] to provide inflationary phase transition. This continuity established with the modification of these parameters result into multi-parameter Van-Der-Waals like EOSs [1-10]. The EOS parameter can provide the necessary conditions for inflation as well as radiation dominated universe, dark matter dominated universe but it fails to discuss the conditions of decelerating phase (after inflation), graceful exit and the reheating phase [1]. The decelerating nature of the expanding universe is the necessary cause to start the late time acceleration of universe and that's why graceful exit from cosmic inflation is necessary [10]. The multi-parameters (viz. $w, \gamma, \rho_c, \alpha$ and $\beta$) models [2-4] or non-linear models can provide the proper cause/conditions for these interior phases. The non-linear fluid can produce negative pressure and also obey strong energy conditions with negative viscosity [11]. Although negative viscosity [11] is quite unnatural in the view of generalized second law of thermodynamics but we can introduce negative viscosity with multi-fluid system. The viscosity has been discussed with a well-established form of coefficient where we have mentioned the results only for two values of power($n = 1$ $and$ $3$) as discussed [8]. The two-different structure has been used to discuss the system under constant and variable viscosity (for $n = 1$ viscosity is constant and for $n = 3$ it is variable). The corresponding scalar field of these non-linear models can discuss the nature of vacuum fields during several cosmological phases. Thus, non-linear fluids can act as a substitute of exotic energy or dark energy in universe evolution [5, 6].

## 2.2. Possible three different formats of non-linear model

Depending upon the choice of $w(\rho, t)$ $and$ $f(\rho)$ [8] we may differentiate the generalized form of that model into three types. They are one parameter model, two parameters model and three parameters models. The details of those models are taken from the works [2, 8] and is summarized in table I. It may be noted, to resolve the issue of dimensions of the denominators of the models i.e. the term $(3 - \rho)$ for one parameter model and $(1 - \beta\rho)$ for three parameters model, we have substituted the following.

$$3 - \rho \rightarrow 3 - \frac{\rho}{\rho_c}$$

And

$$1 - \beta\rho \rightarrow 1 - \beta\frac{\rho}{\rho_c}$$

This approach will not affect our basic focus of study of non-linear equation of state.

**Table I**

| One parameter model | Two parameters model | Three parameters model |
|---|---|---|
| Pressure with viscous effect | Pressure with viscous effect | Pressure with viscous effect |
| $p_m = \dfrac{8w\rho}{3 - \dfrac{\rho}{\rho_c}} - 3\rho^2 - \tau(3H)^{n+1}$ | $p_m = \dfrac{\gamma\rho}{1 - \dfrac{1}{3\rho_c}\rho} - \dfrac{9\gamma}{8\rho_c}\rho^2 - \tau(3H)^{n+1}$ | $p_m = \dfrac{\gamma\rho}{1 - \dfrac{\beta\rho}{\rho_c}} - \alpha\rho^2 - \tau(3H)^{n+1}$ |
| EOS parameter | EOS parameter | EOS parameter |
| $w_{EOS} = \frac{p_m}{\rho} = \frac{8w}{3 - \frac{\rho}{\rho_c}} - 3\rho - \frac{\tau}{\rho}(3H)^{n+1}$ | $w_{EOS} = \frac{p_m}{\rho} = \frac{\gamma}{1 - \frac{1}{3\rho_c}\rho} - \frac{9\gamma}{8\rho_c}\rho - \frac{\tau}{\rho}(3H)^{n+1}$ | $w_{EOS} = \frac{P_m}{\rho} = \frac{\gamma}{1 - \frac{\beta\rho}{\rho_c}} - \alpha\rho - \frac{\tau}{\rho}(3H)^{n+1}$ |
| EOS parameter with $n = 1$ $$w_{EOS} = \frac{8w}{3 - \frac{\rho}{\rho_c}} - 3\rho - 3\tau$$ | EOS parameter with $n = 1$ $$w_{EOS} = \frac{\gamma}{1 - \frac{1}{3\rho_c}\rho} - \frac{9\gamma}{8\rho_c}\rho - 3\tau$$ | EOS parameter with $n = 1$ $$w_{EOS} = \frac{\gamma}{1 - \frac{\beta\rho}{\rho_c}} - \alpha\rho - 3\tau$$ |
| EOS parameter with $n = 3$ $$w_{EOS} = \frac{8w}{3 - \frac{\rho}{\rho_c}} - 3\rho - 9\tau\rho$$ | EOS parameter with $n = 3$ $$w_{EOS} = \frac{\gamma}{1 - \frac{1}{3\rho_c}\rho} - \frac{9\gamma}{8\rho_c}\rho - 9\tau\rho$$ | EOS parameter with $n = 3$ $$w_{EOS} = \frac{\gamma}{1 - \frac{\beta\rho}{\rho_c}} - \alpha\rho - 9\tau\rho$$ |

Here we have assumed for one parameter model $w(t, \rho) = \frac{8w}{3 - \frac{\rho}{\rho_c}}$ where $\frac{C}{b} = 8w, b = \frac{1}{3}$ and $a = 3$; for two parameters model $w(t, \rho) = \frac{\gamma}{1 - \frac{1}{3\rho_c}\rho}$ where $C = \gamma, b = \frac{1}{3\rho_c}$ and $a = \frac{9\gamma}{8\rho_c}$ and for three parameters model $w(t, \rho) = \frac{\gamma}{1 - \frac{\beta\rho}{\rho_c}}$ where $C = \gamma, b = \beta$ and $a = \alpha$. This can be written in terms of critical values of the variables and they are $C = \frac{c_s^2}{c^2}, b = \frac{1}{3}v_c$ and $a = 3p_c v_c^2$. Here $c_s = sound\ speed, c = light\ speed, p_c = critical\ pressure, v_c = critical\ volume$. [5, 10]

## 3. Interior fluid thermodynamics

Now to proceed with our study we need to simplify the above models with our proper targets. Now as already mentioned that we aim to introduce the nonlinear inhomogeneous model [32-34] in the view of Van-der-Waals EOS [2], we have to simplify the above models such that the nonlinearity in them is not destroyed. We will simplify these models just to derive the internal energy as a function of scale factor [12,13]. Here in this paper we have introduced the viscosity just to bring the stability in the models [12, 13]. Therefore, from our assumptions the viscosity will not be an interior property of the cosmic fluid. This is an exterior property that can only modify the pressure [8, 10].Therefore, the energy density will remain unchanged as for without viscous cases. The pressure will be negative due dissipation with viscosity [8]. In the following sections we will show the pressure, energy density, internal energy, EOS parameter (with modified pressure), Temperature and entropy and their change with volume scale factor [12, 13]. All those variables will be plotted with the values of constants for which the unmodified pressure will be positive. Before going into further details, let us have a look at the units of the cosmological parameters to be studied in the subsequent part of the study. The scale factor is dimensionless. For the dark energy density, one may note that the dark energy density is considered to be less than its critical density $\rho_c = 3M_p^2 H^2$, which is approximately $(3 \times 10^{-12} GeV)^4 \approx 10^{-46} GeV^4$. In an expanding universe, the thermodynamic behavior of the dark energy is important. For thermodynamic analysis, the initial CMB temperature is $T_0^{CMB} = 2.73K$. Pressure of dark energy is the product of equation of state parameter and density of the dark energy candidate [58-60].

For simplicity of our calculation we considered critical density $\rho_c = 1$, which helped us in the graphical representation to make the energy density axis to be unit free. (as we have taken the values of energy density w.r.t. the critical density scale, i.e. the variable became $\frac{\rho}{\rho_c}$ instead of only $\rho$). [Appendix]

To plot the graph of the other variables we initialized the values of the variables for $V \to 0$. We plotted the other variables (viz. pressure, Scalar field kinetic energy and Scalar field potential) taking those initial values as unit scale. That's why all those variables became unitless in the plots. (Scalar field kinetic energy and potential energy have been given in section 6) [Appendix]

In the case of Internal energy, we observed the initial value of Internal energy becomes zero for $V \to 0$. So, we scale it in terms of integration constant $U_0$. [Appendix]

### 3.1. One parameter model:

The one parameter model can be considered as follows [2, 3, 4].

$$P = \frac{8w\rho}{3-\frac{\rho}{\rho_c}} - 3\rho^2 \approx \frac{8w}{3}\rho + \left(\frac{8w}{9} - 3\right)\rho^2 \qquad (14)$$

Now considering $p = -\left(\frac{\partial U}{\partial V}\right)$ and $\rho = \frac{U}{V}$ [12, 13] we may write the following differential equation for internal energy. Here U = Internal energy and V = Volume scale factor.$= (a(t))^3$, $a(t)$ is the scale factor of the universe.

By calculation we get

$$U = \frac{(8w+3)V}{(27-8w)+(9+24w)U_0 V^{\frac{8w+3}{3}}} \qquad (15)$$

Where $U_0$ is a positive constant

We get the definition of energy density and pressure [12, 13] as $= \frac{U}{V}$ and $p = -\left(\frac{\partial U}{\partial V}\right)_S$. Then,

$$\rho = \frac{(8w+3)}{(27-8w)+(9+24w)U_0 V^{\frac{8w+3}{3}}} \qquad (16)$$

And the pressure will be as follows,

$$p = \frac{8w\left(\frac{(8w+3)}{(27-8w)+(9+24w)U_0 V^{\frac{8w+3}{3}}}\right)}{3-\left(\frac{(8w+3)}{(27-8w)+(9+24w)U_0 V^{\frac{8w+3}{3}}}\right)} - 3\left(\frac{(8w+3)}{(27-8w)+(9+24w)U_0 V^{\frac{8w+3}{3}}}\right)^2 \tag{17}$$

Thus, we can get the variable pressure and density and therefore the EOS parameter which is as follows. (for n = 1 and 3) [8]

$$w_{EOS} = \frac{P_m}{\rho} = \frac{8w}{3-\left(\frac{(8w+3)}{(27-8w)+(9+24w)U_0 V^{\frac{8w+3}{3}}}\right)} - 3\left(\frac{(8w+3)}{(27-8w)+(9+24w)U_0 V^{\frac{8w+3}{3}}}\right) - 3\tau \tag{18}$$

And

$$w_{EOS} = \frac{P_m}{\rho} = \frac{8w}{3-\left(\frac{(8w+3)}{(27-8w)+(9+24w)U_0 V^{\frac{8w+3}{3}}}\right)} - 3\left(\frac{(8w+3)}{(27-8w)+(9+24w)U_0 V^{\frac{8w+3}{3}}}\right) - 9\tau\left(\frac{(8w+3)}{(27-8w)+(9+24w)U_0 V^{\frac{8w+3}{3}}}\right) \tag{19}$$

Now the temperature will become as follows [12, 13].

$$T = \exp\left(-\int_{V_0}^{V} \frac{dV}{V}\left[\frac{24w}{(3-\rho)^2} - 6\rho\right]\right) \tag{20}$$

And the entropy will be as follows [12, 13].

$$\Delta S = \frac{(8w+3)(9+24w)V^{\frac{8w+6}{3}}}{\left[(27-8w)+(9+24w)V^{\frac{8w+3}{3}}U_0\right]^2} \frac{U_0}{\exp\left(-\int_{V_0}^{V} \frac{dV}{V}\left[\frac{24w}{(3-\rho)^2} - 6\rho\right]\right)} \tag{21}$$

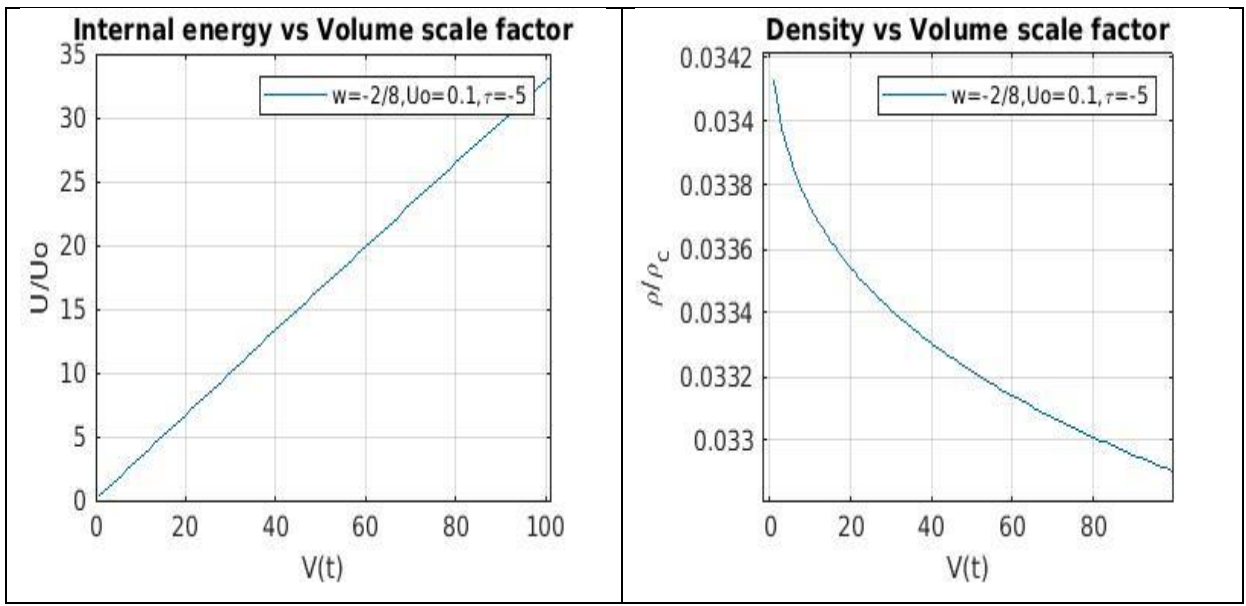

| Figure 1: Graph for internal energy $w = -\frac{2}{8}, U_0 = 0.1$ and $\tau = -5$ | Figure 2: Graph for Energy density($\rho$); $w = -\frac{2}{8}, U_0 = 0.1$ and $\tau = -5$ |
|---|---|
| 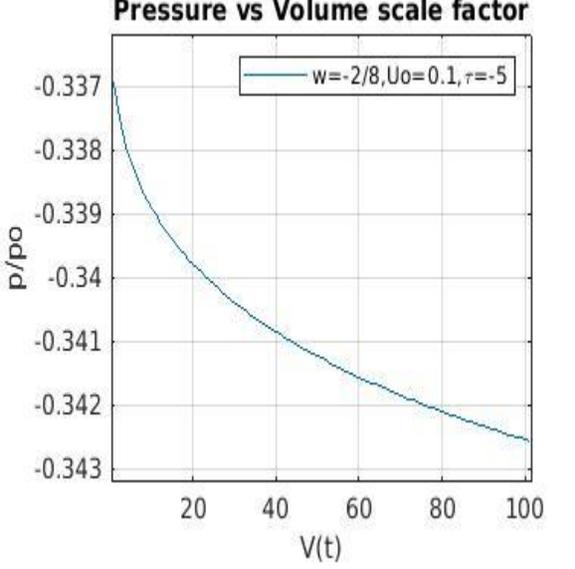 | 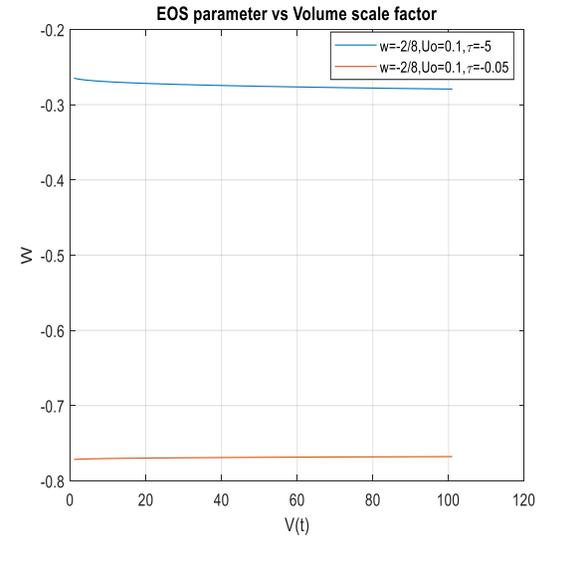 |
| Figure 3: Graph for Pressure (p); $w = -\frac{2}{8}, U_0 = 0.1$ and $\tau = -5$ | Figure 4: Graph for EOS parameter; $w = -\frac{2}{8}, U_0 = 0.1$ and $\tau = -5$; $\tau = -0.05$ |

We have plotted the schematics showing the evolution of internal energy, energy density, pressure and EOS parameters in Figs. 1, 2, 3 and 4 respectively. It may be noted that here we have considered the value of $U_0 > 0$ and viscosity $\tau < 0$. Which state that we have considered the negative viscosity in our fluid system. Here w is also negative in value.

### 3.2. Two parameters model:

The one parameter model can be considered as follows [2, 8, 10].

$$P = \frac{\gamma\rho}{1-\frac{1}{3\rho_c}\rho} - \frac{9\gamma}{8\rho_c}\rho^2 \approx \gamma\rho - \frac{19\gamma}{24\rho_c}\rho^2 \quad (22)$$

Now considering $p = -\left(\frac{\partial U}{\partial V}\right)$ and $\rho = \frac{U}{V}$ [12, 13] we may write the following differential equation for internal energy. Here U = Internal energy and V = Volume scale factor.

We get Internal energy as

$$U = \frac{24(\gamma+1)\rho_c V}{19\gamma + 24 U_0 \rho_c (1+\gamma) V^{1+\gamma}} \quad (23)$$

Here the term $U_0$ is the integration constant which is also a function of entropy only.

So, we get the definition of energy density and pressure as $= \frac{U}{V}$ and $p = -\left(\frac{\partial U}{\partial V}\right)_S$. Then,

$$\rho = \frac{24(\gamma+1)\rho_c}{19\gamma + 24 U_0 \rho_c (1+\gamma) V^{1+\gamma}} \tag{24}$$

And the pressure will be as follows,

$$p = \frac{\gamma\left(\frac{24(\gamma+1)\rho_c}{19\gamma + 24 U_0 \rho_c (1+\gamma) V^{1+\gamma}}\right)}{1 - \frac{1}{3\rho_c}\left(\frac{24(\gamma+1)\rho_c}{19\gamma + 24 U_0 \rho_c (1+\gamma) V^{1+\gamma}}\right)} - \frac{9\gamma}{8\rho_c}\left(\frac{24(\gamma+1)\rho_c}{19\gamma + 24 U_0 \rho_c (1+\gamma) V^{1+\gamma}}\right)^2 \tag{25}$$

Thus, we can get the variable pressure and density and therefore the EOS parameter which is as follows. (for n = 1 and 3) [8]

$$w_{EOS} = \frac{P_m}{\rho} = \frac{\gamma}{1 - \frac{1}{3\rho_c}\left(\frac{24(\gamma+1)\rho_c}{19\gamma + 24 U_0 \rho_c (1+\gamma) V^{1+\gamma}}\right)} - \frac{9\gamma}{8\rho_c}\left(\frac{24(\gamma+1)\rho_c}{19\gamma + 24 U_0 \rho_c (1+\gamma) V^{1+\gamma}}\right) - 3\tau \tag{26}$$

And

$$w_{EOS} = \frac{P_m}{\rho} = \frac{\gamma}{1 - \frac{1}{3\rho_c}\left(\frac{24(\gamma+1)\rho_c}{19\gamma + 24 U_0 \rho_c (1+\gamma) V^{1+\gamma}}\right)} - \frac{9\gamma}{8\rho_c}\left(\frac{24(\gamma+1)\rho_c}{19\gamma + 24 U_0 \rho_c (1+\gamma) V^{1+\gamma}}\right) - 9\tau\left(\frac{24(\gamma+1)\rho_c}{19\gamma + 24 U_0 \rho_c (1+\gamma) V^{1+\gamma}}\right) \tag{27}$$

Now the temperature will become as follows [12, 13].

$$T = \exp\left(-\int_{V_0}^{V} \frac{dV}{V}\left[\frac{\gamma}{\left(3 - \frac{1}{3\rho_c}\rho\right)^2} - \frac{9\gamma}{4\rho_c}\rho\right]\right) \tag{27a}$$

And the entropy will be as follows [12, 13].

$$\Delta S = \frac{[24(\gamma+1)]^2 V^{\gamma+2} \rho_c}{[19\gamma + 24(1+\gamma)\rho_c V^{\gamma+1} U_0]^2} \frac{U_0}{\exp\left(-\int_{V_0}^{V} \frac{dV}{V}\left[\frac{\gamma}{\left(3 - \frac{1}{3\rho_c}\rho\right)^2} - \frac{9\gamma}{4\rho_c}\rho\right]\right)} \tag{27b}$$

| 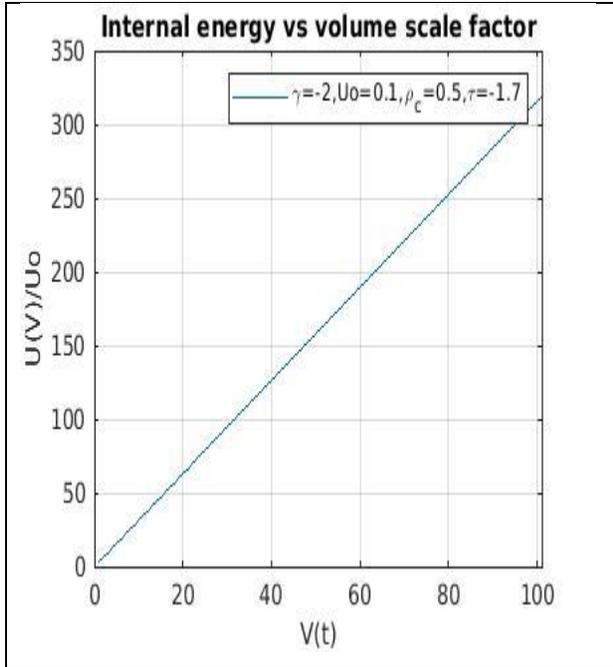 | 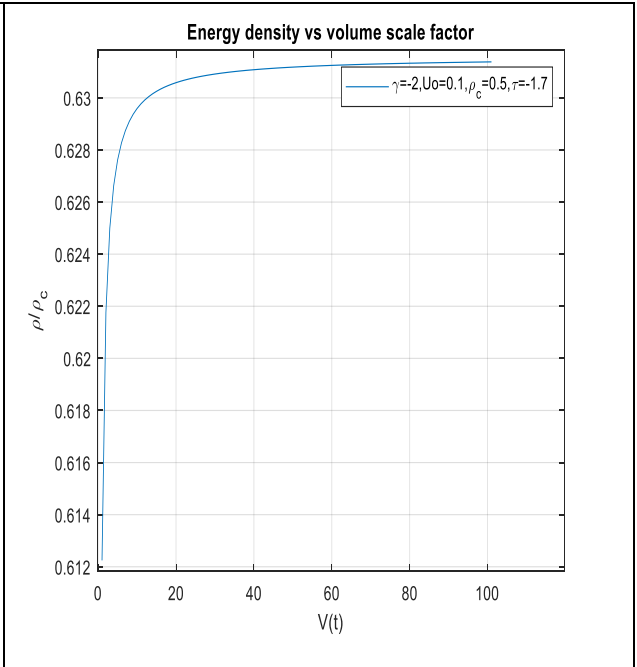 |
| --- | --- |
| Figure 5: Graph for internal energy; $\gamma = -2; U_0 = 0.5;$ $\tau = -1.7$ | Figure 6: Graph for Energy density; $\gamma = -2; U_0 = 0.5;$ $\tau = -1.7$ |
| 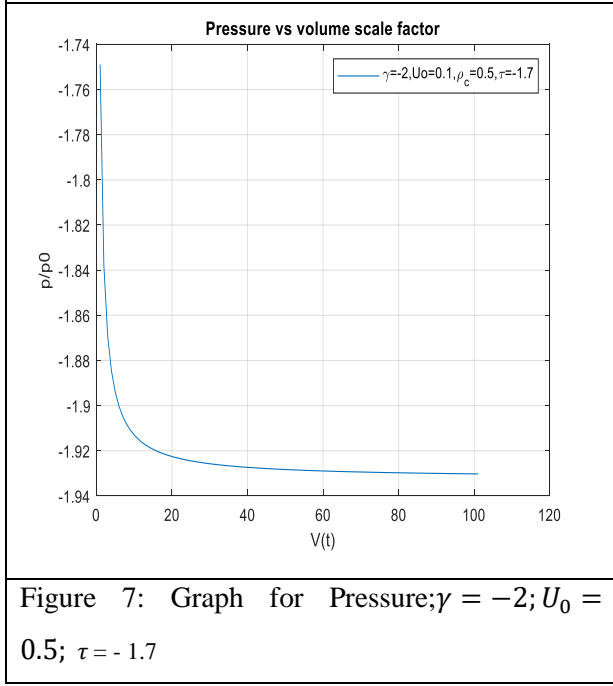 | 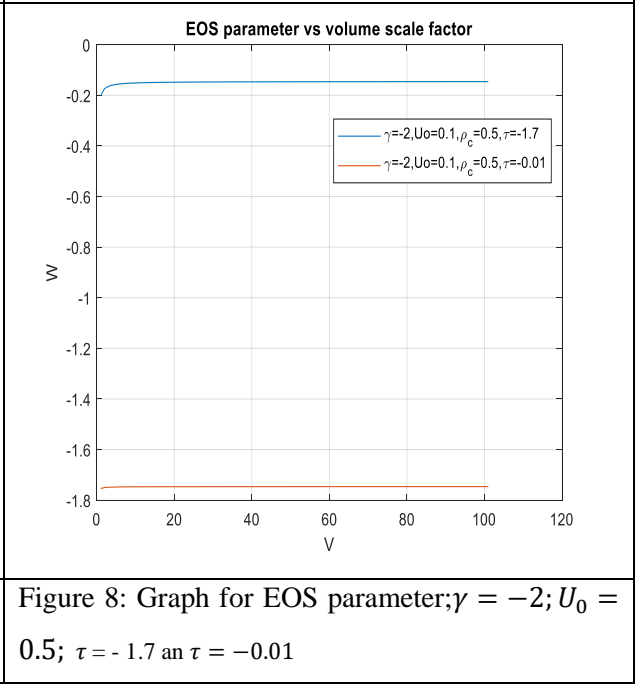 |
| Figure 7: Graph for Pressure; $\gamma = -2; U_0 = 0.5;$ $\tau = -1.7$ | Figure 8: Graph for EOS parameter; $\gamma = -2; U_0 = 0.5;$ $\tau = -1.7$ an $\tau = -0.01$ |

For this model, we have presented graphically the evolution internal energy, energy density, pressure and EOS parameters in Figs. 5, 6, 7 and 8 respectively. Here we have considered the value of $U_0 > 0$ and viscosity $\tau < 0$. Which says that we have considered the negative viscosity in our fluid system. Here $\gamma$ is also negative in value but $\rho_c > 0$.

### 3.3. Three parameters model:

The one parameter model can be considered as follows [2, 9].

$$P = \frac{\gamma \rho}{1 - \frac{\beta \rho}{\rho_c}} - \alpha \rho^2 \approx \gamma \rho + (\gamma \beta - \alpha) \rho^2 \tag{28}$$

Now considering $p = -\left(\frac{\partial U}{\partial V}\right)_S$ and $\rho = \frac{U}{V}$ [12, 13] we may write the following differential equation for internal energy. Here U = Internal energy and V = Volume scale factor.

$$U = \frac{(\gamma+1)V}{(\alpha-\beta\gamma)+U_0(1+\gamma)V^{1+\gamma}} \tag{29}$$

Here the term $U_0$ is the integration constant which is also a function of entropy only.

So, we get the definition of energy density and pressure as $= \frac{U}{V}$ and $p = -\left(\frac{\partial U}{\partial V}\right)_S$. Then,

$$\rho = \frac{(\gamma+1)}{(\alpha-\beta\gamma)+U_0(1+\gamma)V^{1+\gamma}} \tag{30}$$

And the pressure will be as follows,

$$p = \frac{\gamma\left(\frac{(\gamma+1)}{(\alpha-\beta\gamma)+U_0(1+\gamma)V^{1+\gamma}}\right)}{1-\beta\left(\frac{(\gamma+1)}{(\alpha-\beta\gamma)+U_0(1+\gamma)V^{1+\gamma}}\right)} - \alpha\left(\frac{(\gamma+1)}{(\alpha-\beta\gamma)+U_0(1+\gamma)V^{1+\gamma}}\right)^2 \tag{31}$$

Thus, we can get the variable pressure and density and therefore the EOS parameter which is as follows. (for n = 1 and 3) [8]

$$w_{EOS} = \frac{P_m}{\rho} = \frac{\gamma}{1-\beta\left(\frac{(\gamma+1)}{(\alpha-\beta\gamma)+U_0(1+\gamma)V^{1+\gamma}}\right)} - \alpha\left(\frac{(\gamma+1)}{(\alpha-\beta\gamma)+U_0(1+\gamma)V^{1+\gamma}}\right) - 3\tau \tag{32}$$

And

$$w_{EOS} = \frac{P_m}{\rho} = \frac{\gamma}{1-\beta\left(\frac{(\gamma+1)}{(\alpha-\beta\gamma)+U_0(1+\gamma)V^{1+\gamma}}\right)} - \alpha\left(\frac{(\gamma+1)}{(\alpha-\beta\gamma)+U_0(1+\gamma)V^{1+\gamma}}\right) - 9\tau\left(\frac{(\gamma+1)}{(\alpha-\beta\gamma)+U_0(1+\gamma)V^{1+\gamma}}\right) \tag{33}$$

And the temperature will be [12, 13],

$$T = \exp\left(-\int_{V_0}^{V}\frac{dV}{V}\left[\frac{\gamma}{(1-\beta\rho)^2} - 2\alpha\rho\right]\right) \tag{33a}$$

And the entropy will be [12, 13],

$$\Delta S = \frac{(\gamma+1)^2 V^{\gamma+2}}{[(\alpha-\beta\gamma)+(\gamma+1)V^{\gamma+1}U_0]^2} \frac{U_0}{\exp\left(-\int_{V_0}^{V}\frac{dV}{V}\left[\frac{\gamma}{(1-\beta\rho)^2}-2\alpha\rho\right]\right)} \tag{33b}$$

| 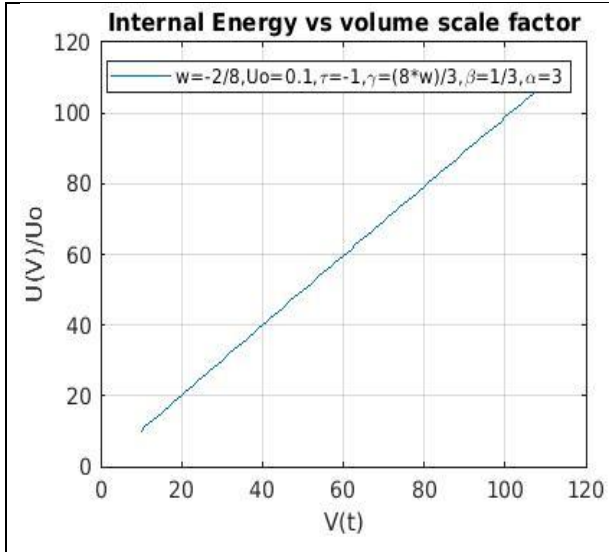 | 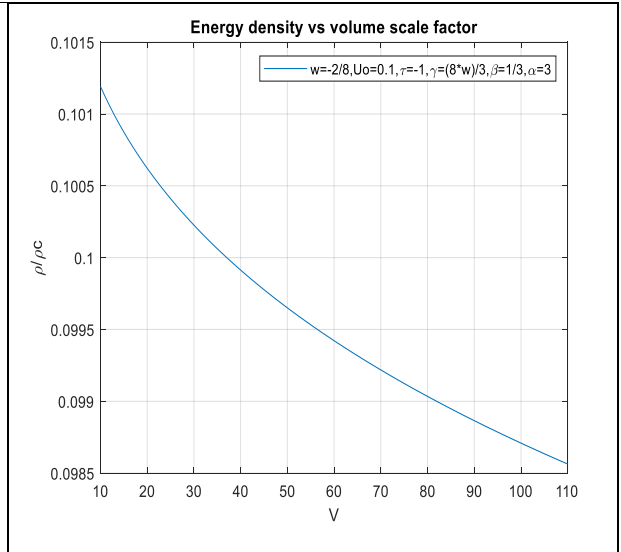 |
|---|---|
| Figure 9: Graph for internal energy $w = -\frac{2}{8}$; $U_0 = 0.1$; $\tau = -1$; $\gamma = \frac{8w}{3}$; $\beta = \frac{1}{3}$; $\alpha = 3$ | Figure 10: Graph for Energy density; $w = -\frac{2}{8}$; $U_0 = 0.1$; $\tau = -1$; $\gamma = \frac{8w}{3}$; $\beta = \frac{1}{3}$; $\alpha = 3$ |
| 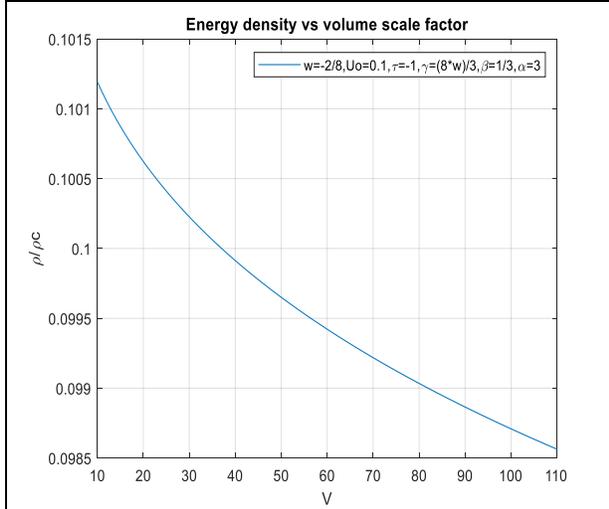 | 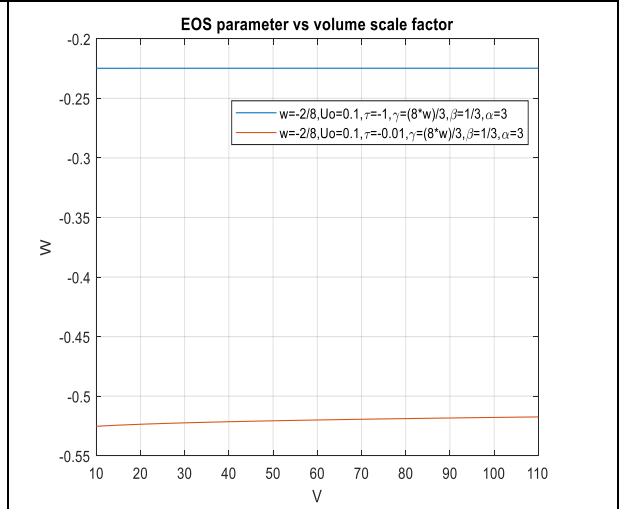 |
| Figure 11: Graph for Pressure; $w = -\frac{2}{8}$; $U_0 = 0.1$; $\tau = -1$; $\gamma = \frac{8w}{3}$; $\beta = \frac{1}{3}$; $\alpha = 3$ | Figure 12: Graph for EOS parameter; $w = -\frac{2}{8}$; $U_0 = 0.1$; $\tau = -1 \, and -0.01$; $\gamma = \frac{8w}{3}$; $\beta = \frac{1}{3}$; $\alpha = 3$ |

Considering the value of $U_0 > 0$ and viscosity $\tau < 0$ we have plotted the internal energy, energy density, pressure and EOS parameter in Figs. 9, 10, 11 and 12 respectively. It is clear that we have considered the negative viscosity in our fluid system. Here w is also negative in value. As in three parameters model $\gamma$ is related with w as $\gamma = \frac{8w}{3}$ so, $\gamma < 0$. The values of $\alpha = 3$ $and$ $\beta = \frac{1}{3}$. For the considered value of constants and parameters we get positive change of entropy.

## 4. Basics of finite time singularity problems and the Thermodynamics Energy conditions:

For finite time singularity problems, we know the following types [15-20, 32];

- Type I (known as Big Rip): At any finite time, $t = t_s$ we will have a $\to \infty$; $\rho \to \infty$; $p \to \infty$
- Type II (known as Sudden Singularity): At any finite time, $t = t_s$ we will have a $\to a_s$; $\rho \to \rho_s$; $p \to \infty$
- Type III: At any finite time, $t = t_s$ we will have a $\to a_s$; $\rho \to \infty$; $p \to \infty$ ; it happens only in the EOS type of $p = -\rho - A\rho^\alpha$.
- Type IV: At any finite time, $t = t_s$ we will have a $\to a_s$; $\rho \to \rho_s$; $p \to p_s$ Moreover, the Hubble rate and its first derivative also remain finite, but the higher derivatives, or some of these diverge. This kind of singularity mainly comes into play when $p = -\rho - f(\rho)$.

The thermodynamic energy conditions are basically derived from the well-known Raychaudhuri's equation. For a congruence of time-like and null-like geodesics, the Raychaudhuri equations are given in the following forms [28];

$$\frac{d\theta}{d\tau} = -\frac{1}{3}\theta^2 - \sigma_{\mu\nu}\sigma^{\mu\nu} + \omega_{\mu\nu}\omega^{\mu\nu} - R_{\mu\nu}u^\mu u^\nu \qquad (34)$$

And

$$\frac{d\theta}{d\tau} = -\frac{1}{3}\theta^2 - \sigma_{\mu\nu}\sigma^{\mu\nu} + \omega_{\mu\nu}\omega^{\mu\nu} - R_{\mu\nu}n^\mu n^\nu \qquad (35)$$

Where $\theta$ is the expansion factor, $n^\mu n^\nu$ is the null vector, and $\sigma_{\mu\nu}\sigma^{\mu\nu}$ and $\omega_{\mu\nu}\omega^{\mu\nu}$ are, respectively, the shear and the rotation associated with the vector field $u^\mu u^\nu$. For attractive gravity we will have the followings;

$$R_{\mu\nu}u^\mu u^\nu \geq 0 \text{ and } R_{\mu\nu}n^\mu n^\nu \geq 0$$

To set some nomenclature the energy conditions of general relativity to be considered here are [28];

(i) Null energy condition (NEC)
(ii) Weak energy condition (WEC)
(iii) Strong energy condition (SEC)
(iv) Dominant energy condition (DEC)

So, for our matter-fluid distribution we may write this condition as follows [21-32];

- NEC = $\rho + p \geq 0$
- WEC = $\rho \geq 0$ and $\rho + p \geq 0$
- SEC = $\rho + 3p \geq 0$ and $\rho + p \geq 0$
- DEC = $\rho \geq 0$ and $-\rho \leq p \leq \rho$

## 5. Resolution of Finite time future singularity problems and analysis of Thermodynamics energy conditions for having attractive gravity with the Van-der-Waals fluid energy density and Pressure:

We will discuss the resolution of Finite time future singularity problem [15-20, 32] with the above derived models. From the above models we will also show that the energy densities and pressures will satisfy the Thermodynamics energy conditions to represent attractive gravity [21-32]. Here we will use the modified pressure as the viscosity is not the interior property of our fluid but our viscosity will help to obey all the energy conditions. Hence the thermodynamics energy conditions will be satisfied with modified pressure of cosmic fluid. We use viscosity for n = 3. [8]

### 5.1. One parameter model:

In this model we found the pressure and density as follows [2-4].

$$p_m = \frac{8w\left(\frac{(8w+3)}{(27-8w)+(9+24w)U_0 V^{\frac{8w+3}{3}}}\right)}{3-\left(\frac{(8w+3)}{(27-8w)+(9+24w)U_0 V^{\frac{8w+3}{3}}}\right)} - 3\left(\frac{(8w+3)}{(27-8w)+(9+24w)U_0 V^{\frac{8w+3}{3}}}\right)^2 - 9\tau\left(\frac{(8w+3)}{(27-8w)+(9+24w)U_0 V^{\frac{8w+3}{3}}}\right)^2$$

(36)

And

$$\rho = \frac{(8w+3)}{(27-8w)+(9+24w)U_0 V^{\frac{8w+3}{3}}} \tag{37}$$

With the above expressions of pressure and density, we study the energy conditions and we get the following graphs (fig. 13 to fig. 15).

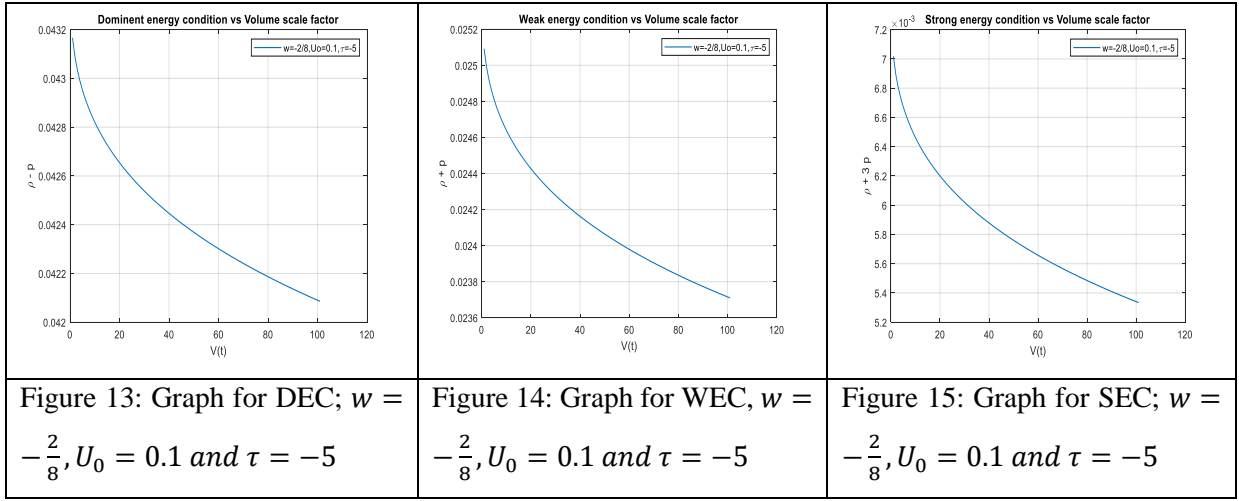

| Figure 13: Graph for DEC; $w = -\frac{2}{8}, U_0 = 0.1$ and $\tau = -5$ | Figure 14: Graph for WEC, $w = -\frac{2}{8}, U_0 = 0.1$ and $\tau = -5$ | Figure 15: Graph for SEC; $w = -\frac{2}{8}, U_0 = 0.1$ and $\tau = -5$ |

For the one parameter model, the DEC, WEC and NEC are plotted in Figs. 13, 14 and 15 and all the conditions are satisfied.

### 5.2. Two parameters model:

In this model we found the pressure and density as follows [2, 8, 10].

$$p_m = \frac{\gamma\left(\frac{24(\gamma+1)\rho_c}{19\gamma+24U_0\rho_c(1+\gamma)V^{1+\gamma}}\right)}{1-\frac{1}{3\rho_c}\left(\frac{24(\gamma+1)\rho_c}{19\gamma+24U_0\rho_c(1+\gamma)V^{1+\gamma}}\right)} - \frac{9\gamma}{8\rho_c}\left(\frac{24(\gamma+1)\rho_c}{19\gamma+24U_0\rho_c(1+\gamma)V^{1+\gamma}}\right)^2 - 9\tau\left(\frac{24(\gamma+1)\rho_c}{19\gamma+24U_0\rho_c(1+\gamma)V^{1+\gamma}}\right)^2$$

(41)

And

$$\rho = \frac{24(\gamma+1)\rho_c}{19\gamma+24U_0\rho_c(1+\gamma)V^{1+\gamma}} \tag{42}$$

For two parameter model, the energy conditions are graphically expressed in fig. 16 to fig. 18.

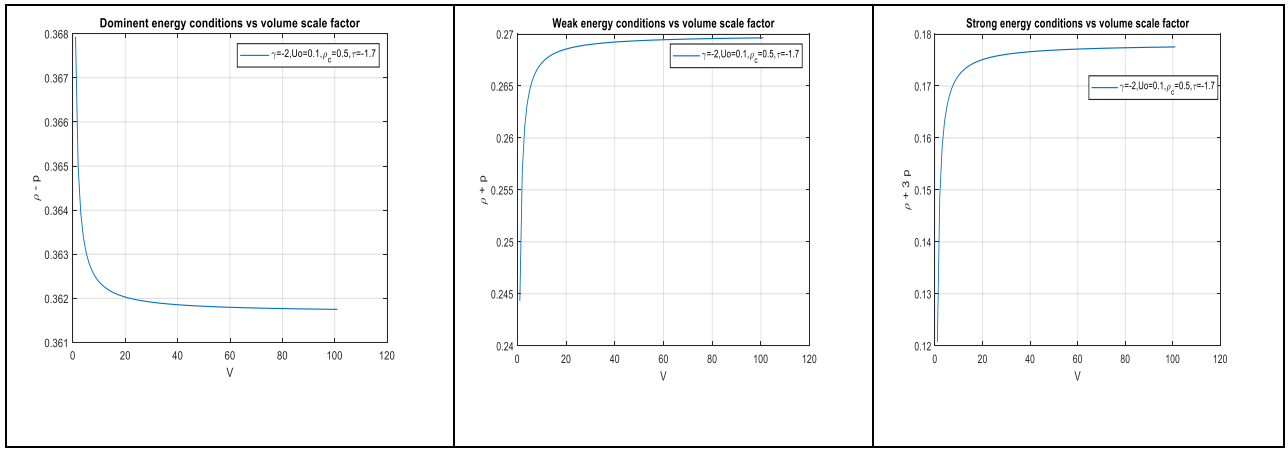

| Figure 16: Graph for DEC; $\gamma = -2; U_0 = 0.5; \tau = -1.7$ | Figure 17: Graph for WEC; $\gamma = -2; U_0 = 0.5; \tau = -1.7$ | Figure 18: Graph for SEC for $\gamma = -2; U_0 = 0.5; \tau = -1.7$ |
|---|---|---|

For the two parameters model, the DEC, WEC and NEC are plotted in Figs. 16, 17 and 18 and all conditions are satisfied.

### 5.3. Three parameters model:

In this model we found the pressure and density as follows [2, 9].

$$p_m = \frac{\gamma\left(\frac{(\gamma+1)}{(\alpha-\beta\gamma)+U_0(1+\gamma)V^{1+\gamma}}\right)}{1-\beta\left(\frac{(\gamma+1)}{(\alpha-\beta\gamma)+U_0(1+\gamma)V^{1+\gamma}}\right)} - \alpha\left(\frac{(\gamma+1)}{(\alpha-\beta\gamma)+U_0(1+\gamma)V^{1+\gamma}}\right)^2 - 9\tau\left(\frac{(\gamma+1)}{(\alpha-\beta\gamma)+U_0(1+\gamma)V^{1+\gamma}}\right)^2 \quad (46)$$

And

$$\rho = \frac{(\gamma+1)}{(\alpha-\beta\gamma)+U_0(1+\gamma)V^{1+\gamma}} \quad (47)$$

For three parameter model, the energy conditions are shown in fig. 19 to fig. 21.

| Figure 19: Graph for DEC; $w = -\frac{2}{8}; U_0 = 0.1; \tau = -1; \gamma = \frac{8w}{3}; \beta = \frac{1}{3}; \alpha = 3$ | Figure 20: Graph for WEC; $w = -\frac{2}{8}; U_0 = 0.1; \tau = -1; \gamma = \frac{8w}{3}; \beta = \frac{1}{3}; \alpha = 3$ | Figure 21: Graph for SEC; $w = -\frac{2}{8}; U_0 = 0.1; \tau = -1; \gamma = \frac{8w}{3}; \beta = \frac{1}{3}; \alpha = 3$ |
|---|---|---|

Therefore, we can observe that the one parameter, two parameters model and three parameters models satisfy all energy conditions including SEC i.e. all the conditions for attractive gravity are obeyed by those three models. Those conditions have been discussed with the previously considered values of constants.

We will discuss the resolution of finite time future singularity problems [15-20] for Van-der-Waals EOS. We consider a minimal coupling between Quintessence scalar field with Van-der-Waals fluid. [32]

$$3H^2 = \rho_\phi + \rho \qquad (51a)$$

Where $\rho_\phi = scalar\ field\ density\ and\ \rho = Van-der-Waals\ fluid\ density$

The scalar field energy density in terms of volume scale factor $V(t) = (a(t))^3$; can be written in the following way from the Klein Gordon equation

$$\rho_\phi = mV^n \qquad (51b)$$

And

The generalized form of fluid density for all those three models can be written as follows.

$$\rho = \frac{A}{B+CV^D} \qquad (51c)$$

Here A, B, C and D are the constants that depend upon the parameters of those three models.

So, from the above three relations we may write as follows.

$$\int \frac{dV}{V}\left[\frac{B+CV^D}{\frac{m}{3}BV^n+\frac{m}{3}CV^{n+D}+\frac{A}{3}}\right] = t + C_1 \qquad (51d)$$

Where $C_1 = constant\ of\ integration \to +ve$

From equation (51d) we can conclude that there is no singularity at $t \to 0\ and\ t \to t_s$ for positive value of $C_1$ (constant). Thus, we have resolved both initial and finite time singularities.

## 6. Corresponding scalar field potential and the kinetic term of scalar field theory:

In this section we will show the corresponding scalar field and potential variation for Van-Der-Waals fluid cosmology [5, 6]. We will use the general scalar field theory or Quintessence to find the scale factor variation of scalar field and its potential. So, the calculations are as follows.

We know from Quintessence scalar field theory [43-52],

$$\rho = \frac{1}{2}\dot\phi^2 + V(\phi)\ and\ p = \frac{1}{2}\dot\phi^2 - V(\phi) \qquad (52)$$

So, we get the kinetic term of scalar field theory and potential as follows.

$$\dot\phi^2 = p + \rho$$

$$\text{Or, } \dot{\phi} = (p + \rho)^{\frac{1}{2}} \tag{53}$$

And

$$V(\phi) = \frac{1}{2}(\rho - p) \tag{54}$$

Now this potential and kinetic term will get the form for the models as follows in table II.

**Table II**

| One parameter model | Two parameters model | Three parameters model |
|---|---|---|
| Scalar field kinetic energy $\dot{\phi}$ $$= \left( 8w \left( \frac{(8w+3)}{(27-8w)+(9+24w)U_0 V^{\frac{8w+3}{3}}} \right) \middle/ \left( 3 - \left( \frac{(8w+3)}{(27-8w)+(9+24w)U_0 V^{\frac{8w+3}{3}}} \right) \right) \right.$$ $$- 3\left( \frac{(8w+3)}{(27-8w)+(9+24w)U_0 V^{\frac{8w+3}{3}}} \right)^2$$ $$\left. + \frac{(8w+3)}{(27-8w)+(9+24w)U_0 V^{\frac{8w+3}{3}}} \right)^{\frac{1}{2}}$$ | Scalar field kinetic energy $\dot{\phi}$ $$= \left( \frac{\gamma \left( \frac{24(\gamma+1)\rho_c}{19\gamma + 24 U_0 \rho_c (1+\gamma) V^{1+\gamma}} \right)}{1 - \frac{1}{3\rho_c}\left( \frac{24(\gamma+1)\rho_c}{19\gamma + 24 U_0 \rho_c (1+\gamma) V^{1+\gamma}} \right)} \right.$$ $$- \frac{9\gamma}{8\rho_c}\left( \frac{24(\gamma+1)\rho_c}{19\gamma + 24 U_0 \rho_c (1+\gamma) V^{1+\gamma}} \right)^2$$ $$\left. + \frac{24(\gamma+1)\rho_c}{19\gamma + 24 U_0 \rho_c (1+\gamma) V^{1+\gamma}} \right)^{\frac{1}{2}}$$ | Scalar field kinetic energy $\dot{\phi}$ $$= \left( \frac{\gamma \left( \frac{(\gamma+1)}{(\alpha - \beta\gamma) + U_0(1+\gamma)V^{1+\gamma}} \right)}{1 - \beta \left( \frac{(\gamma+1)}{(\alpha - \beta\gamma) + U_0(1+\gamma)V^{1+\gamma}} \right)} \right.$$ $$- \alpha \left( \frac{(\gamma+1)}{(\alpha - \beta\gamma) + U_0(1+\gamma)V^{1+\gamma}} \right)^2$$ $$\left. + \frac{(\gamma+1)}{(\alpha - \beta\gamma) + U_0(1+\gamma)V^{1+\gamma}} \right)^{\frac{1}{2}}$$ |
| Scalar field potential $V(\phi)$ $$= \frac{1}{2}\left( \frac{(8w+3)}{(27-8w)+(9+24w)U_0 V^{\frac{8w+3}{3}}} \right.$$ $$- \frac{8w\left( \frac{(8w+3)}{(27-8w)+(9+24w)U_0 V^{\frac{8w+3}{3}}} \right)}{3 - \left( \frac{(8w+3)}{(27-8w)+(9+24w)U_0 V^{\frac{8w+3}{3}}} \right)}$$ $$\left. + 3\left( \frac{(8w+3)}{(27-8w)+(9+24w)U_0 V^{\frac{8w+3}{3}}} \right)^2 \right)$$ | Scalar field potential $V(\phi)$ $$= \frac{1}{2}\left( \frac{24(\gamma+1)\rho_c}{19\gamma + 24 U_0 \rho_c (1+\gamma) V^{1+\gamma}} \right.$$ $$- \frac{\gamma\left( \frac{24(\gamma+1)\rho_c}{19\gamma + 24 U_0 \rho_c (1+\gamma) V^{1+\gamma}} \right)}{1 - \frac{1}{3\rho_c}\left( \frac{24(\gamma+1)\rho_c}{19\gamma + 24 U_0 \rho_c (1+\gamma) V^{1+\gamma}} \right)}$$ $$\left. + \frac{9\gamma}{8\rho_c}\left( \frac{24(\gamma+1)\rho_c}{19\gamma + 24 U_0 \rho_c (1+\gamma) V^{1+\gamma}} \right)^2 \right)$$ | Scalar field potential $V(\phi)$ $$= \frac{1}{2}\left( \frac{(\gamma+1)}{(\alpha - \beta\gamma) + U_0(1+\gamma)V^{1+\gamma}} \right.$$ $$- \frac{\gamma\left( \frac{(\gamma+1)}{(\alpha - \beta\gamma) + U_0(1+\gamma)V^{1+\gamma}} \right)}{1 - \beta\left( \frac{(\gamma+1)}{(\alpha - \beta\gamma) + U_0(1+\gamma)V^{1+\gamma}} \right)}$$ $$\left. + \alpha\left( \frac{(\gamma+1)}{(\alpha - \beta\gamma) + U_0(1+\gamma)V^{1+\gamma}} \right)^2 \right)$$ |

Now the scalar field kinetic energy and potential energies are plotted in fig. 22 to 27.

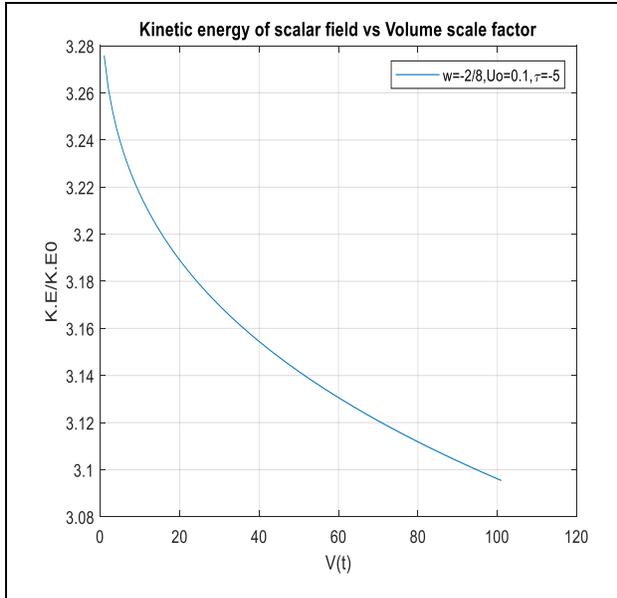

Figure 22: Graph for Kinetic term in fields; $w = -\frac{2}{8}, U_0 = 0.1\ and\ \tau = -5$

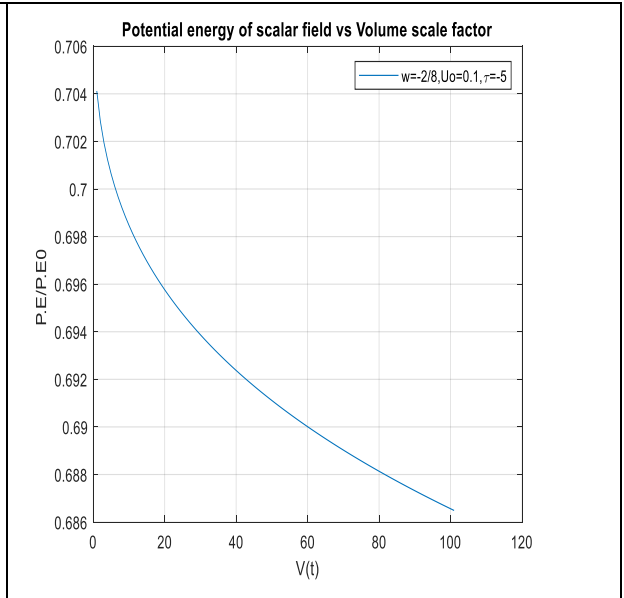

Figure 23: Graph for potential term in fields; $w = -\frac{2}{8}, U_0 = 0.1\ and\ \tau = -5$

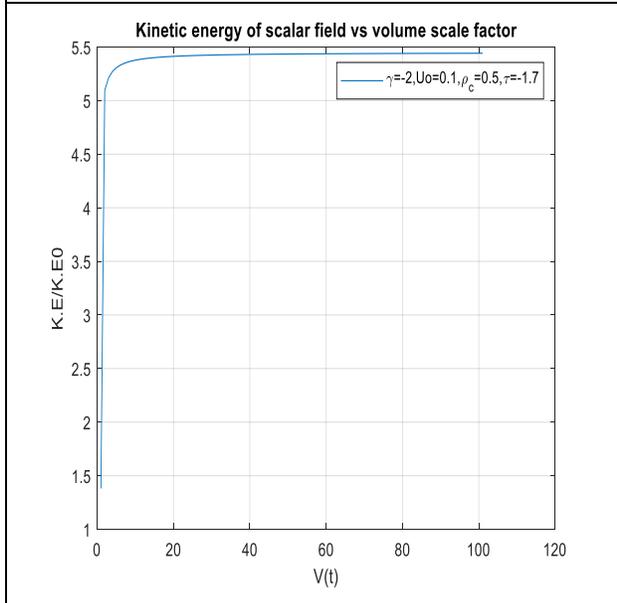

Figure 24: Graph for Kinetic term in fields with modified pressure for $\gamma = -2; U_0 = 0.5;\ \tau = -1.7$

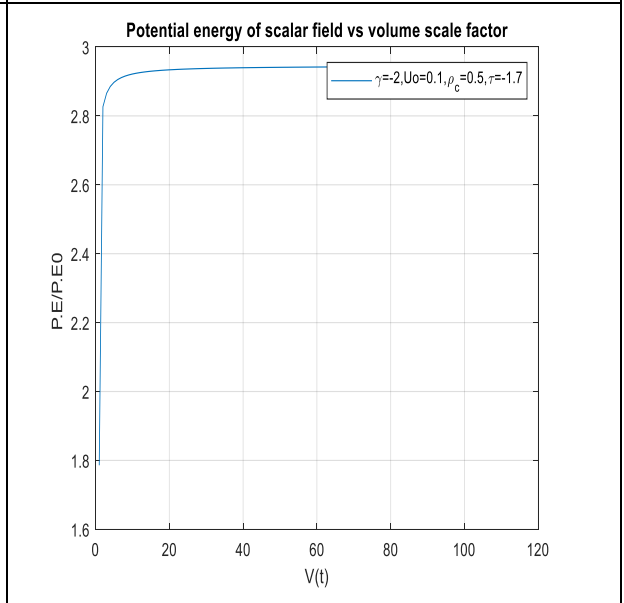

Figure 25: Graph for potential term in fields with modified pressure for $\gamma = -2; U_0 = 0.5;\ \tau = -1.7$

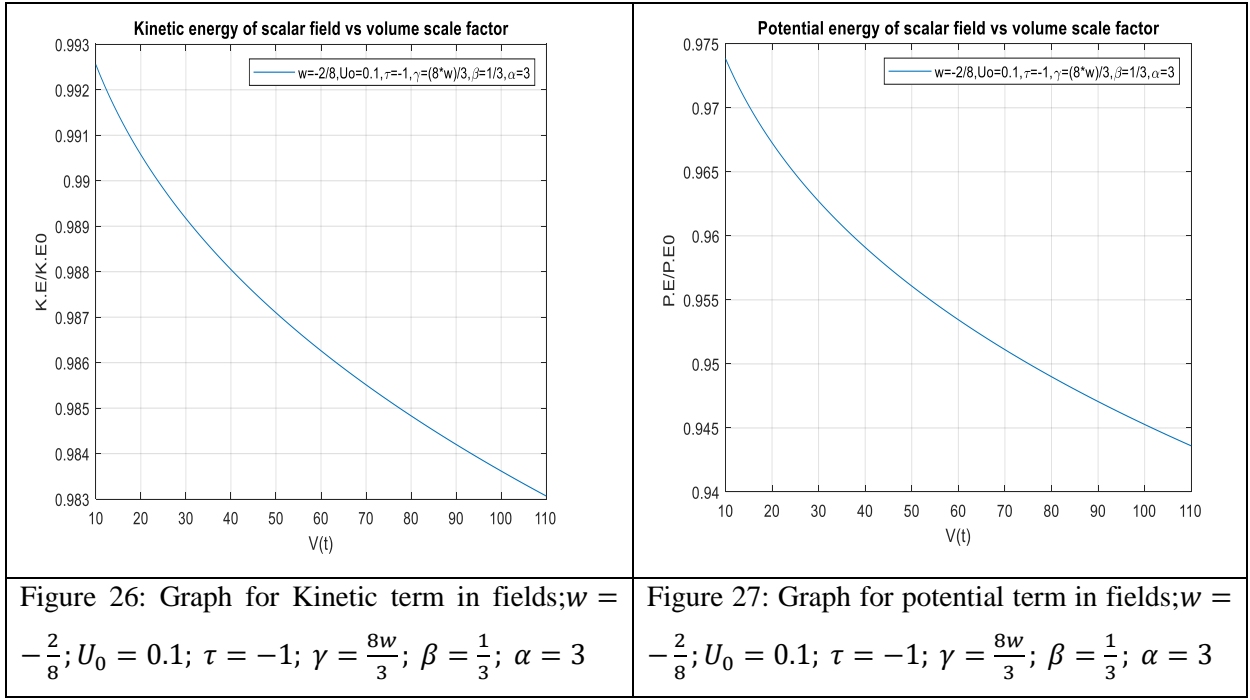

| Figure 26: Graph for Kinetic term in fields; $w = -\frac{2}{8}; U_0 = 0.1; \tau = -1; \gamma = \frac{8w}{3}; \beta = \frac{1}{3}; \alpha = 3$ | Figure 27: Graph for potential term in fields; $w = -\frac{2}{8}; U_0 = 0.1; \tau = -1; \gamma = \frac{8w}{3}; \beta = \frac{1}{3}; \alpha = 3$ |
|---|---|

We have discussed the scalar field potential and the kinetic energy of scalar field. We can observe that for all the three models with earlier considered values of constants and parameters, the potential has much higher value than the kinetic energy which is a criteria of late time acceleration and inflation [5, 6]. Again, the potentials have decreasing nature which also provides the proof of slow roll mechanism [43]. Therefore, for the negative viscosity we can explain slow roll scalar field mechanism properly with all those three models.

## 7. Thermodynamics stability analysis for this fluid for unmodified pressure:

Here we will now discuss the thermodynamics stability of those three models with respect to expanding universe. As we know that the expanding universe fluid pressure and internal energy should satisfy the following three rules of model stability. In this section all the calculations have been done by following the papers of Panigrahi, D. and Chatterjee, S [12, 13]. Calculations have been shown only for unmodified pressures.

$$\text{Rule I: } \left(\frac{\partial p}{\partial V}\right)_S < 0$$

$$\text{Rule II: } \left(\frac{\partial p}{\partial V}\right)_T < 0$$

$$\text{Rule III: } c_V = T\left(\frac{\partial S}{\partial T}\right)_V > 0$$

According those rules of stability for those three models may be listed as follows.

## 7.1. One parameter model:

From Rule I: Here we have used the rule I for model stability.

$$\left(\frac{\partial p}{\partial V}\right)_S = -\left[\frac{24w}{(3-\rho)^2} - 6\rho\right]\frac{U_0(8w+3)^2(9+24w)V^{\frac{8w}{3}}}{3\left[(27-8w)+(9+24w)V^{\frac{8w+3}{3}}U_0\right]^2} \tag{61}$$

From Rule II: Here we have used the rule II for model stability.

$$\left(\frac{\partial p}{\partial V}\right)_T = -\left[\frac{24w}{(3-\rho)^2} - 6\rho\right]\left[\left(\frac{(8w+3)ST}{U_0(9+24w)}\right)^{\frac{1}{2}}\left(\frac{8w+6}{6}\right)V^{-\frac{8w+12}{6}} - \frac{1}{2}\left(\frac{(8w+3)TS^{-1}V^{-\frac{(8w+6)}{3}}}{U_0(9+24w)}\right)^{\frac{1}{2}}\frac{(8w+3)(8w+6)(9+24w)U_0V^{\frac{8w+9}{3}}}{3T\left[(27-8w)+(9+24w)V^{\frac{8w+3}{3}}U_0\right]^2} +$$

$$\frac{1}{2}\left(\frac{(8w+3)TS^{-1}V^{-\frac{(8w+6)}{3}}}{U_0(9+24w)}\right)^{\frac{1}{2}}\frac{2(8w+3)^2(9+24w)^2U_0^2V^{\frac{16w+12}{3}}}{T\left[(27-8w)+(9+24w)V^{\frac{8w+3}{3}}U_0\right]^3}\right] \tag{62}$$

From Rule III: Here we have used the rule III for model stability.

$$c_V = T\left(\frac{\partial S}{\partial T}\right)_V = -\frac{(8w+3)(9+24w)V^{\frac{8w+6}{3}}}{\left[(27-8w)+(9+24w)V^{\frac{8w+3}{3}}U_0\right]^2}\frac{U_0}{T} \tag{63}$$

## 7.2. Two parameters model:

From Rule I: Here we have used the rule I for model stability.

$$\left(\frac{\partial p}{\partial V}\right)_S = -\left[\frac{\gamma}{\left(3-\frac{1}{3\rho_c}\rho\right)^2} - \frac{9\gamma}{4\rho_c}\rho\right]\frac{U_0(24)^2(\gamma+1)^3\rho_cV^\gamma}{[19\gamma+24\rho_c(\gamma+1)V^{(\gamma+1)}U_0]^2} \tag{64}$$

From Rule II: Here we have used the rule II for model stability.

$$\left(\frac{\partial p}{\partial V}\right)_T = -\left[\frac{\gamma}{\left(3-\frac{1}{3\rho_c}\rho\right)^2} - \frac{9\gamma}{4\rho_c}\rho\right]\left[\left(\frac{\rho_cST}{U_0}\right)^{\frac{1}{2}}\left(\frac{\gamma+2}{2}\right)V^{-\frac{\gamma+4}{2}}\right.$$

$$-\frac{1}{2}\left(\frac{\rho_cTS^{-1}V^{-(\gamma+2)}}{cU_0}\right)^{\frac{1}{2}}\frac{(24(\gamma+1))^2\rho_c(\gamma+2)U_0V^{\gamma+1}}{T[19\gamma+24\rho_c(\gamma+1)V^{(\gamma+1)}U_0]^2}$$

$$\left.+\frac{1}{2}\left(\frac{\rho_cTS^{-1}V^{-(\gamma+2)}}{U_0}\right)^{\frac{1}{2}}\frac{2(24(\gamma+1))^3\rho_c^2U_0^2V^{2\gamma+2}}{T[19\gamma+24\rho_c(\gamma+1)V^{(\gamma+1)}U_0]^3}\right]$$

$$\tag{65}$$

From Rule III: Here we have used the rule III for model stability.

$$c_V = T\left(\frac{\partial S}{\partial T}\right)_V = -\frac{(24(\gamma+1))^2 \rho_c V^{\gamma+2}}{[19\gamma + 24\rho_c(\gamma+1)V^{(\gamma+1)}U_0]^2}\frac{U_0}{T} \tag{66}$$

### 7.3. Three parameters model:

From Rule I: Here we have used the rule I for model stability.

$$\left(\frac{\partial p}{\partial V}\right)_S = -\left[\frac{\gamma}{(1-\beta\rho)^2} - 2\alpha\rho\right]\frac{U_0(\gamma+1)^3 V^\gamma}{[(\alpha-\beta\gamma)+(\gamma+1)V^{(\gamma+1)}U_0]^2} \tag{67}$$

From Rule II: Here we have used the rule II for model stability.

$$\left(\frac{\partial p}{\partial V}\right)_T = -\left[\frac{\gamma}{(1-\beta\rho)^2} - 2\alpha\rho\right]\left[\left(\frac{ST}{U_0}\right)^{\frac{1}{2}}\left(\frac{\gamma+2}{2}\right)V^{-\frac{\gamma+4}{2}} - \frac{1}{2}\left(\frac{TS^{-1}V^{-(\gamma+2)}}{U_0}\right)^{\frac{1}{2}}\frac{(\gamma+1)^2(\gamma+2)U_0 V^{\gamma+1}}{T[(\alpha-\beta\gamma)+(\gamma+1)V^{(\gamma+1)}U_0]^2} + \frac{1}{2}\left(\frac{TS^{-1}V^{-(\gamma+2)}}{U_0}\right)^{\frac{1}{2}}\frac{(\gamma+1)^4 U_0^2 V^{2\gamma+2}}{T[(\alpha-\beta\gamma)+(\gamma+1)V^{(\gamma+1)}U_0]^3}\right] \tag{68}$$

From Rule III: Here we have used the rule III for model stability.

$$c_V = T\left(\frac{\partial S}{\partial T}\right)_V = -\frac{(\gamma+1)^2 V^{\gamma+2}}{[(\alpha-\beta\gamma)+(\gamma+1)V^{(\gamma+1)}U_0]^2}\frac{U_0}{T} \tag{69}$$

Here in this section we can observe that for without viscosity case the stability conditions can't be satisfied with those Van-der-Waals models. The Rule I and rule II are the conditions that can't be obeyed by the models. That's why in the proceeding section we will discuss the same with negative viscosity.

## 8. Thermodynamics stability with modified pressure:

Here we will now discuss the thermodynamics stability of those three models with respect to expanding universe for viscous Van-der-Waals fluid. As we know that the expanding universe fluid pressure and internal energy should satisfy the earlier said three rules of model stability. We will show the results for both n = 1 and 3 [8]. [12, 13] Here the calculations for modified pressures have been discussed.

### 8.1. One parameter model:

From Rule I: Here we have used the rule I for model stability with modified pressures.

$$\left(\frac{\partial p_m}{\partial V}\right)_S = -\left[\frac{24w}{(3-\rho)^2} - 6\rho - 3\tau\right]A \tag{70}$$

And for n = 3

$$\left(\frac{\partial p_m}{\partial V}\right)_S = -\left[\frac{24w}{(3-\rho)^2} - 6\rho - 6\tau\rho\right]A \qquad (71)$$

Here $A = \dfrac{U_0(8w+3)^2(9+24w)V^{\frac{8w}{3}}}{3\left[(27-8w)+(9+24w)V^{\frac{8w+3}{3}}U_0\right]^2}$

From Rule II: Here we have used the rule II for model stability with modified pressures.

$$\left(\frac{\partial p_m}{\partial V}\right)_T = -\left[\frac{24w}{(3-\rho)^2} - 6\rho - 3\tau\right]B \qquad (72)$$

And for n = 3

$$\left(\frac{\partial p_m}{\partial V}\right)_T = -\left[\frac{24w}{(3-\rho)^2} - 6\rho - 6\tau\rho\right]B \qquad (73)$$

Here
$$B = \left[\left(\frac{(8w+3)ST}{U_0(9+24w)}\right)^{\frac{1}{2}}\left(\frac{8w+6}{6}\right)V^{-\frac{8w+12}{6}} - \frac{1}{2}\left(\frac{(8w+3)TS^{-1}V^{-\frac{(8w+6)}{3}}}{U_0(9+24w)}\right)^{\frac{1}{2}}\frac{(8w+3)(8w+6)(9+24w)U_0V^{\frac{8w+9}{3}}}{3T\left[(27-8w)+(9+24w)V^{\frac{8w+3}{3}}U_0\right]^2} + \right.$$

$$\left. \frac{1}{2}\left(\frac{(8w+3)TS^{-1}V^{-\frac{(8w+6)}{3}}}{U_0(9+24w)}\right)^{\frac{1}{2}}\frac{2(8w+3)^2(9+24w)^2U_0^2V^{\frac{16w+12}{3}}}{T\left[(27-8w)+(9+24w)V^{\frac{8w+3}{3}}U_0\right]^3}\right]$$

From Rule III: Here we have used the rule III for model stability with modified pressures.

$$c_V = T\left(\frac{\partial S}{\partial T}\right)_V = -\frac{(8w+3)(9+24w)V^{\frac{8w+6}{3}}}{\left[(27-8w)+(9+24w)V^{\frac{8w+3}{3}}U_0\right]^2}\frac{U_0}{T} \qquad (74)$$

## 8.2. Two parameters model:

From Rule I: Here we have used the rule I for model stability with modified pressures.

$$\left(\frac{\partial p_m}{\partial V}\right)_S = -\left[\frac{\gamma}{\left(3-\frac{1}{3\rho_c}\rho\right)^2} - \frac{9\gamma}{4\rho_c}\rho - 3\tau\right]C \qquad (75)$$

And for n = 3

$$\left(\frac{\partial p_m}{\partial V}\right)_S = -\left[\frac{\gamma}{\left(3-\frac{1}{3\rho_c}\rho\right)^2} - \frac{9\gamma}{4\rho_c}\rho - 6\tau\rho\right]C \qquad (76)$$

Here $C = \dfrac{U_0(24)^2(\gamma+1)^3\rho_c V^\gamma}{\left[19\gamma + 24\rho_c(\gamma+1)V^{(\gamma+1)}U_0\right]^2}$

From Rule II: Here we have used the rule II for model stability with modified pressures.

$$\left(\frac{\partial p_m}{\partial V}\right)_T = -\left[\frac{\gamma}{\left(3-\frac{1}{3\rho_c}\rho\right)^2} - \frac{9\gamma}{4\rho_c}\rho - 3\tau\right]D \tag{77}$$

And for n = 3

$$\left(\frac{\partial p_m}{\partial V}\right)_T = -\left[\frac{\gamma}{\left(3-\frac{1}{3\rho_c}\rho\right)^2} - \frac{9\gamma}{4\rho_c}\rho - 6\tau\rho\right]D \tag{78}$$

Here $D = \left[\left(\frac{\rho_c ST}{U_0}\right)^{\frac{1}{2}}\left(\frac{\gamma+2}{2}\right)V^{-\frac{\gamma+4}{2}} - \frac{1}{2}\left(\frac{\rho_c TS^{-1}V^{-(\gamma+2)}}{cU_0}\right)^{\frac{1}{2}}\frac{(24(\gamma+1))^2\rho_c(\gamma+2)U_0V^{\gamma+1}}{T[19\gamma+24\rho_c(\gamma+1)V^{(\gamma+1)}U_0]^2} +$

$\frac{1}{2}\left(\frac{\rho_c TS^{-1}V^{-(\gamma+2)}}{U_0}\right)^{\frac{1}{2}}\frac{2(24(\gamma+1))^3\rho_c^2 U_0^2 V^{2\gamma+2}}{T[19\gamma+24\rho_c(\gamma+1)V^{(\gamma+1)}U_0]^3}\right]$

From Rule III: Here we have used the rule III for model stability with modified pressures.

$$c_V = T\left(\frac{\partial S}{\partial T}\right)_V = -\frac{(24(\gamma+1))^2\rho_c V^{\gamma+2}}{[19\gamma+24\rho_c(\gamma+1)V^{(\gamma+1)}U_0]^2}\frac{U_0}{T} \tag{79}$$

**8.3. Three parameters model:**

From Rule I: Here we have used the rule I for model stability with modified pressures.

$$\left(\frac{\partial p_m}{\partial V}\right)_S = -\left[\frac{\gamma}{(1-\beta\rho)^2} - 2\alpha\rho - 3\tau\right]E \tag{80}$$

And for n = 3

$$\left(\frac{\partial p_m}{\partial V}\right)_S = -\left[\frac{\gamma}{(1-\beta\rho)^2} - 2\alpha\rho - 6\tau\rho\right]E \tag{81}$$

Here $E = \frac{U_0(\gamma+1)^3 V^\gamma}{[(\alpha-\beta\gamma)+(\gamma+1)V^{(\gamma+1)}U_0]^2}$

From Rule II: Here we have used the rule II for model stability with modified pressures.

$$\left(\frac{\partial p_m}{\partial V}\right)_T = -\left[\frac{\gamma}{(1-\beta\rho)^2} - 2\alpha\rho - 3\tau\right]F \tag{82}$$

And for n = 3

$$\left(\frac{\partial p_m}{\partial V}\right)_T = -\left[\frac{\gamma}{(1-\beta\rho)^2} - 2\alpha\rho - 6\tau\rho\right]F \qquad (83)$$

Here
$$F = \left[\left(\frac{ST}{U_0}\right)^{\frac{1}{2}}\left(\frac{\gamma+2}{2}\right)V^{-\frac{\gamma+4}{2}} - \frac{1}{2}\left(\frac{TS^{-1}V^{-(\gamma+2)}}{U_0}\right)^{\frac{1}{2}}\frac{(\gamma+1)^2(\gamma+2)U_0 V^{\gamma+1}}{T[(\alpha-\beta\gamma)+(\gamma+1)V^{(\gamma+1)}U_0]^2} + \frac{1}{2}\left(\frac{TS^{-1}V^{-(\gamma+2)}}{U_0}\right)^{\frac{1}{2}}\frac{(\gamma+1)^4 U_0^2 V^{2\gamma+2}}{T[(\alpha-\beta\gamma)+(\gamma+1)V^{(\gamma+1)}U_0]^3}\right]$$

From Rule III: Here we have used the rule III for model stability with modified pressures.

$$c_V = T\left(\frac{\partial S}{\partial T}\right)_V = -\frac{(\gamma+1)^2 V^{\gamma+2}}{[(\alpha-\beta\gamma)+(\gamma+1)V^{(\gamma+1)}U_0]^2}\frac{U_0}{T} \qquad (84)$$

All the rules for stability conditions are satisfied for the considered values of constant. Here we have introduced the negative viscosity with the modification of pressure. Thus, we can observe that even rule I and Rule II have been satisfied with all those three Van-Der-Waals models.

## 9. Concluding Remarks:

In this work, we have investigated the thermodynamics and energy conditions of cosmological fluid described by Van der Waals equation of state with viscosity. Viscosity acts as a dissipating factor. For one, two and three parameter models, the modified pressure comes out to be negative while internal energy and energy density came out to be positive. Introduction of viscosity has made all the models thermodynamically stable. But the only problem here comes out to be negative heat capacity. So, to get positive heat capacity we must introduce the idea of multiple fluid system, so that the summation of heat capacity of all the fluids come out to be positive.

During the introduction of viscosity, we have used the negative coefficient of viscosity that helped us in satisfying all the stability conditions as well as thermodynamics energy conditions with proper inflationary representations. All those three models have shown similar results after using this kind of new viscous definition of our cosmic fluid. Although we know that negative viscosity for a single type of fluid is not viable in thermodynamics. So, we may conclude that our results in this paper provide strict predictions about the existence of multiple fluid mechanism in cosmic evolution. On the other-hand we have found some idea of multiple fluid during the discussion of heat capacity. The negative heat capacity is only possible when the system itself will reduce energy or degrees of freedom. In other word we may say that the negative heat capacity provides the proof that the Van-Der-Waals system must transfer its energy to other system. So, we may again conclude that the negative viscosity increases the fluid energy and the negative heat capacity decreases the fluid energy, which keep the system energy conserved. Thus, the Van-der-Waals fluid models are capable of explaining the cosmic evolution only when we consider multiple fluid system. [ special ref. is in appendix equation (A10)]

In the plots of EOS parameter, we observe that Van-der-Waals fluid can also discuss the phantom era perfectly including the resolution of Finite time future singularity problems. The plots of scalar field potentials and scalar field kinetic energy provide the proof of slow roll mechanism as well as negative pressure to provide the accelerating expansion of universe. This outcome is in consistency with the study of Brevik et al. [37], where inflationary expansion was described in terms of a Van-der -Waals equation of state for the cosmic fluid in presence of bulk viscosity. Furthermore, From equation (51d) (three parameters model) we have observed that there is no singularity at $t \to 0$ $and$ $t \to t_s$ for positive value of $C_1$ (constant). Thus, both initial and finite time singularities have been resolved under the current model under study. Therefore, with all those results and the values of parameters and constants we have now solved the thermodynamics stability problems, finite time future singularity problems, inflation and negative pressure problem, initial singularity problem with satisfying the thermodynamics energy conditions and energy conservation principles for attractive gravity.

While concluding, let us comment on the possibility of viscous Little-Rip singularity under the current cosmological models, whose EOS parameters are given in Eqs. (19), (27) and (33). The EOS parameters are plotted in Figs. 4, 8 and 12. In this connection, let us consider the work of Brevik et al. [49], where the authors considered the role of a viscous (or inhomogeneous (imperfect) equation of state) fluid in a Little Rip cosmology. In Little Rip cosmology, EOS asymptotically tends to -1 from <-1 and thus avoids singularity. In the present cases, the EOS has not shown any asymptotic behavior. Thus, under the current cosmological framework, the presence of viscosity is not found to Little Rip singularity.

Our main aim in this work was to explain the accelerated expansion of the universe and simultaneous satisfaction of thermodynamics energy conditions with the help of non-linear equation of state which in this case is adopted in Van-der-Waals form. Our model proved the presence of negative pressure which is strictly necessary for the accelerated expansion of universe. Hence, we can say that the Van-der-Waals equation of state provides the similar results as obtained from dark energy models. Experimental evidences of accelerated universe support our obtained results. [62-65]

**Appendix:**

From two fluid interacting system we may write as follows.

$$\dot{\rho}_1 + 3H(\rho_1 + p_1) = -Q_1 \tag{A1}$$

$$\dot{\rho}_2 + 3H(\rho_2 + p_2) = Q_1 \tag{A2}$$

Introducing a term $\Delta$ where $Q_1 = 3H\Delta$ and $\Delta = 3\eta(t)(3H)^n$ we can write as follows.

$$\dot{\rho}_1 + 3H(\rho_1 + p_{1m}) = 0 \tag{A3}$$

$$\dot{\rho}_2 + 3H(\rho_2 + p_{2m}) = 0 \tag{A4}$$

Where $p_{1m} = p_1 + 3\eta(t)(3H)^n$ and $p_{2m} = p_2 - 3\eta(t)(3H)^n$. So, for the system which has negative viscosity will gain energy.

Now we get,

$$c_V = T\left(\frac{\partial S}{\partial T}\right)_V = \left(\frac{\partial U}{\partial T}\right)_V$$

$$\text{Or, } \Delta U = c_V \Delta T = Q_2 = 3H\Delta_1 \tag{A5}$$

So, we can again write as follows.

$$\dot{\rho}_1 + 3H(\rho_1 + p_{1m}) = Q_2 \tag{A6}$$

$$\dot{\rho}_2 + 3H(\rho_2 + p_{2m}) = -Q_2 \tag{A7}$$

Or we may write this as follows.

$$\dot{\rho}_1 + 3H(\rho_1 + p'_{1m}) = 0 \tag{A8}$$

$$\dot{\rho}_2 + 3H(\rho_2 + p'_{2m}) = 0 \tag{A9}$$

Where $p'_{1m} = p_{1m} - \Delta_1 = p_1 + 3\eta(t)(3H)^n - \Delta_1$ and $p'_{2m} = p_{2m} + \Delta_1 = p_1 - 3\eta(t)(3H)^n + \Delta_1$

Thus, the pressure will be modified with the energy transition between the multiple fluid systems. Now to satisfy the second law of thermodynamics we need to conserve the system energy. So, we must conclude that $\Delta_1 = 3\eta(t)(3H)^n$. Therefore, we may write again as follows.

$$c_V \Delta T = Q_2 = 3H\Delta_1 = 9H\eta(t)(3H)^n \tag{A10}$$

From the equation (A10) we can observe the direct relation between heat capacity and energy transition for viscosity.

The variables $\rho$, $p$, $U$ has been calculated analytically and we have tried to study the nature of their evolution. We have derived those values w.r.t. their initial values so that the variables become unitless in the graphical representations. As we haven't used any pre-established critical values of those variables, we never used the terms density parameter, pressure parameter etc. in our work.

**Acknowledgement:** The authors are thankful to the anonymous reviewer for their insightful comments. Financial support under the CSIR Grant No. 03(1420)/18/EMRII is thankfully acknowledged by Surajit Chattopadhyay.

**Reference:**

1. Kremer, G.M., 2004. Brane cosmology with a van der Waals equation of state. *General Relativity and Gravitation*, *36*(6), pp.1423-1432.
2. Vardiashvili, G., Halstead, E., Poltis, R., Morgan, A. and Tobar, D., 2017. Inflationary constraints on the van der Waals equation of state. *arXiv preprint arXiv:1701.00748*.
3. Jantsch, R.C., Christmann, M.H. and Kremer, G.M., 2016. The van der Waals fluid and its role in cosmology. *International Journal of Modern Physics D*, *25*(03), p.1650031.
4. Kremer, G.M., 2003. Cosmological models described by a mixture of van der Waals fluid and dark energy. *Physical Review D*, *68*(12), p.123507.
5. Capozziello, S., De Martino, S. and Falanga, M., 2002. Van der Waals quintessence. *Physics Letters A*, *299*(5-6), pp.494-498.
6. Capozziello, S., Carloni, S. and Troisi, A., 2003. Quintessence without scalar fields. *arXiv preprint astro-ph/0303041*.
7. Ivanov, R.I. and Prodanov, E.M., 2019. Van der Waals universe with adiabatic matter creation. *The European Physical Journal C*, *79*(2), p.118.
8. Brevik, I., Obukhov, V.V. and Timoshkin, A.V., 2018. Inflation in terms of a viscous van der Waals coupled fluid. *International Journal of Geometric Methods in Modern Physics*, *15*(09), p.1850150.
9. Elizalde, E. and Khurshudyan, M., 2018. Cosmology with an interacting van der Waals fluid. *International Journal of Modern Physics D*, *27*(04), p.1850037.
10. Obukhov, V.V. and Timoshkin, A.V., 2018. Cosmological Van Der Waals Model with Viscosity in an Inflationary Universe. *Russian Physics Journal*, *60*(10), pp.1705-1711.
11. Brevik, I. and Grøn, Ø., 2013. Universe models with negative bulk viscosity. *Astrophysics and Space Science*, *347*(2), pp.399-404.
12. Panigrahi, D. and Chatterjee, S., 2017. Viability of variable generalised Chaplygin gas: a thermodynamical approach. *General Relativity and Gravitation*, *49*(3), p.35.
13. Panigrahi, D. and Chatterjee, S., 2016. Thermodynamics of the variable modified Chaplygin gas. *Journal of Cosmology and Astroparticle Physics*, *2016*(05), p.052.
14. Chakraborty, S., Guha, S. and Panigrahi, D., 2019. Evolution of FRW universe in variable modified Chaplygin gas model. *arXiv preprint arXiv:1906.12185*.

15. Brevik, I., Timoshkin, A.V. and Paul, T., 2021. The effect of thermal radiation on singularities in the dark universe. *arXiv preprint arXiv:2103.08430*.
16. Odintsov, S.D. and Oikonomou, V.K., 2018. Dynamical systems perspective of cosmological finite-time singularities in f (R) gravity and interacting multifluid cosmology. *Physical Review D*, *98*(2), p.024013.
17. Odintsov, S.D. and Oikonomou, V.K., 2017. Big bounce with finite-time singularity: The F (R) gravity description. *International Journal of Modern Physics D*, *26*(08), p.1750085.
18. Frampton, P.H., Ludwick, K.J. and Scherrer, R.J., 2012. Pseudo-rip: Cosmological models intermediate between the cosmological constant and the little rip. *Physical Review D*, *85*(8), p.083001.
19. Frampton, P.H., Ludwick, K.J., Nojiri, S., Odintsov, S.D. and Scherrer, R.J., 2012. Models for little rip dark energy. *Physics Letters B*, *708*(1-2), pp.204-211.
20. Frampton, P.H., Ludwick, K.J. and Scherrer, R.J., 2011. The little rip. *Physical Review D*, *84*(6), p.063003.
21. Visser, M. and Barcelo, C., 2000. Energy conditions and their cosmological implications. In *Cosmo-99* (pp. 98-112).
22. Chattopadhyay, S., Pasqua, A. and Khurshudyan, M., 2014. New holographic reconstruction of scalar-field dark-energy models in the framework of chameleon Brans–Dicke cosmology. *The European Physical Journal C*, *74*(9), pp.1-13.
23. Arora, S., Santos, J.R.L. and Sahoo, P.K., 2021. Constraining f (Q, T) gravity from energy conditions. *Physics of the Dark Universe*, *31*, p.100790.
24. Sharma, U.K. and Pradhan, A., 2018. Cosmology in modified f (R, T)-gravity theory in a variant Λ (T) scenario-revisited. *International Journal of Geometric Methods in Modern Physics*, *15*(01), p.1850014.
25. Sahoo, P.K., Mandal, S. and Arora, S., 2021. Energy conditions in non-minimally coupled f (R, T) gravity. *Astronomische Nachrichten*, *342*(1-2), pp.89-95.
26. Yadav, A.K., Sahoo, P.K. and Bhardwaj, V., 2019. Bulk viscous Bianchi-I embedded cosmological model in f (R, T)= f 1 (R)+ f 2 (R) f 3 (T) gravity. *Modern Physics Letters A*, *34*(19), p.1950145.
27. Sharma, L.K., Singh, B.K. and Yadav, A.K., 2020. Viability of Bianchi type V universe in f (R, T)= f 1 (R)+ f 2 (R) f 3 (T) gravity. *International Journal of Geometric Methods in Modern Physics*, *17*(07), p.2050111.
28. Moraes, P.H.R.S. and Sahoo, P.K., 2017. The simplest non-minimal matter–geometry coupling in the f (R, T) cosmology. *The European Physical Journal C*, *77*(7), pp.1-8.

29. Hulke, N., Singh, G.P., Bishi, B.K. and Singh, A., 2020. Variable Chaplygin gas cosmologies in f (R, T) gravity with particle creation. *New Astronomy*, *77*, p.101357.
30. Singla, N., Gupta, M.K. and Yadav, A.K., 2020. Accelerating Model of a Flat Universe in $\boldsymbol{f(R, T)}$ Gravity. *Gravitation and Cosmology*, *26*(2), pp.144-152.
31. Sharif, M., Rani, S. and Myrzakulov, R., 2013. Analysis of F (R, T) gravity models through energy conditions. *The European Physical Journal Plus*, *128*(10), pp.1-11.
32. Kar, A., Sadhukhan, S. and Chattopadhyay, S., 2021. Energy conditions for inhomogeneous EOS and its thermodynamics analysis with the resolution on finite time future singularity problems. *International Journal of Geometric Methods in Modern Physics*, Vol. 18, No. 08, 2150131 (2021)
33. Chakraborty, W. and Debnath, U., 2008. Interaction between scalar field and ideal fluid with inhomogeneous equation of state. *Physics Letters B*, *661*(1), pp.1-4.
34. Nojiri, S.I. and Odintsov, S.D., 2005. Inhomogeneous equation of state of the universe: Phantom era, future singularity, and crossing the phantom barrier. *Physical Review D*, *72*(2), p.023003.
35. Nojiri, S.I. and Odintsov, S.D., 2006. The new form of the equation of state for dark energy fluid and accelerating universe. *Physics Letters B*, *639*(3-4), pp.144-150.
36. Khadekar, G.S. and Raut, D., 2018. FRW viscous fluid cosmological model with time-dependent inhomogeneous equation of state. *International Journal of Geometric Methods in Modern Physics*, *15*(01), p.1830001.
37. Štefančić, H., 2009. The solution of the cosmological constant problem from the inhomogeneous equation of state—a hint from modified gravity?. *Physics Letters B*, *670*(4-5), pp.246-253.
38. Brevik, I., Elizalde, E., Gorbunova, O. and Timoshkin, A.V., 2007. A FRW dark fluid with a non-linear inhomogeneous equation of state. *The European Physical Journal C*, *52*(1), pp.223-228.
39. Myrzakulov, R., Sebastiani, L. and Zerbini, S., 2013. Inhomogeneous Viscous Fluids in a Friedmann-Robertson-Walker (FRW) Universe. *Galaxies*, *1*(2), pp.83-95.
40. Jamil, M. and Rashid, M.A., 2008. Interacting dark energy with inhomogeneous equation of state. *The European Physical Journal C*, *56*(3), pp.429-434.
41. Khadekar, G.S., Raut, D. and Miskin, V.G., 2015. FRW viscous cosmology with inhomogeneous equation of state and future singularity. *Modern Physics Letters A*, *30*(29), p.1550144.
42. Varshney, G., Sharma, U.K., Pradhan, A. and Kumar, N., 2021. Reconstruction of Tachyon, Dirac-Born-Infeld-essence and Phantom model for Tsallis holographic dark energy in f (R, T) gravity. *Chinese Journal of Physics*.
43. Tsujikawa, S., 2013. Quintessence: a review. *Classical and Quantum Gravity*, *30*(21), p.214003.

44. Banerjee, A. and Koley, R., 2016. Inflationary field excursion in broad classes of scalar field models. *Physical Review D*, *94*(12), p.123506.
45. Hughes, J., 2019. THE QUINTESSENTIAL DARK ENERGY THEORY: QUINTESSENCE.
46. Steinhardt, P.J., 2003. A quintessential introduction to dark energy. *Philosophical Transactions of the Royal Society of London. Series A: Mathematical, Physical and Engineering Sciences*, *361*(1812), pp.2497-2513.
47. Capozziello, S., Carloni, S. and Troisi, A., 2003. Quintessence without scalar fields. *arXiv preprint astro-ph/0303041*.
48. Kremer, G.M., 2003. Cosmological models described by a mixture of van der Waals fluid and dark energy. *Physical Review D*, *68*(12), p.123507.
49. Elizalde, E. and Khurshudyan, M., 2018. Cosmology with an interacting van der Waals fluid. *International Journal of Modern Physics D*, *27*(04), p.1850037.
50. Sadhukhan. S, Quintessence Model Calculations for Bulk Viscous Fluid and Low Value Predictions of the Coefficient of Bulk Viscosity, International Journal of Science and Research (IJSR) 9(3):1419-1420, DOI: 10.21275/SR20327132301
51. Kar. A, Sadhukhan. S ,HAMILTONIAN FORMALISM FOR BIANCHI TYPE I MODEL FOR PERFECT FLUID AS WELL AS FOR THE FLUID WITH BULK AND SHEARING VISCOSITY, Basic and Applied Sciences into Next Frontiers, ISBN: 978-81-948993-0-3; New Delhi Publishers
52. Kar. A, Sadhukhan. S , QUINTESSENCE MODEL WITH BULK VISCOSITY AND SOME PREDICTIONS ON THE COEFFICIENT OF BULK VISCOSITY AND GRAVITATIONAL CONSTANT, RECENT ADVANCEMENT OF MATHEMATICS IN SCIENCE AND TECHNOLOGY (ISBN): 978-81-950475-0-5, RECENT ADVANCEMENT OF MATHEMATICS IN SCIENCE AND TECHNOLOGY (ISBN): 978-81-950475-0-5
53. Sinha, S., Chattopadhyay, S. and Radinschi, I., 2019. Cosmology of viscous holographic f (G) gravity and consequences in the framework of quintessence scalar field. *International Journal of Geometric Methods in Modern Physics*, *16*(11), p.1950176.
54. Chattopadhyay, S. and Karmakar, S., 2019. Reconstruction of f (T) gravity in the context of standard Chaplygin gas as tachyon scalar field and study of the stability against gravitational perturbation. *International Journal of Geometric Methods in Modern Physics*, *16*(07), p.1950101.
55. Chattopadhyay, S., 2017. Interacting modified Chaplygin gas in f (T) gravity framework and analysis of its stability against gravitational perturbation. *International Journal of Geometric Methods in Modern Physics*, *14*(03), p.1750035.

56. Chattopadhyay, S., 2017. Modified Chaplygin gas equation of state on viscous dissipative extended holographic Ricci dark energy and the cosmological consequences. *International Journal of Modern Physics D*, *26*(06), p.1750042.
57. Chattopadhyay, S., 2017. A study on the bouncing behavior of modified Chaplygin gas in presence of bulk viscosity and its consequences in the modified gravity framework. *International Journal of Geometric Methods in Modern Physics*, *14*(12), p.1750181.
58. Karmakar, S., Chattopadhyay, S. and Radinschi, I., 2020. A holographic reconstruction scheme for f (R) gravity and the study of stability and thermodynamic consequences. *New Astronomy*, *76*, p.101321.
59. M. Li, X-D. Li, S. Wang, Y. Wang, Dark Energy, ISSN: 2382-5960, World Scientific.
60. Cárdenas, V.H., Grandón, D. and Lepe, S., 2019. Dark energy and dark matter interaction in light of the second law of thermodynamics. The European Physical Journal C, 79(4), pp.1-10.
61. Arjona, R., 2020. Machine learning meets the redshift evolution of the CMB temperature. Journal of Cosmology and Astroparticle Physics, 2020(08), p.009.
62. Perlmutter, S., Aldering, G., Della Valle, M., Deustua, S., Ellis, R.S., Fabbro, S., Fruchter, A., Goldhaber, G., Groom, D.E., Hook, I.M. and Kim, A.G., 1998. Discovery of a supernova explosion at half the age of the Universe. *Nature*, *391*(6662), pp.51-54.
63. Riess, A.G., Filippenko, A.V., Challis, P., Clocchiatti, A., Diercks, A., Garnavich, P.M., Gilliland, R.L., Hogan, C.J., Jha, S., Kirshner, R.P. and Leibundgut, B.R.U.N.O., 1998. Observational evidence from supernovae for an accelerating universe and a cosmological constant. *The Astronomical Journal*, *116*(3), p.1009.
64. Perlmutter, S., Aldering, G., Goldhaber, G., Knop, R.A., Nugent, P., Castro, P.G., Deustua, S., Fabbro, S., Goobar, A., Groom, D.E. and Hook, I.M., 1999. Measurements of Ω and Λ from 42 high-redshift supernovae. *The Astrophysical Journal*, *517*(2), p.565.
65. Schmidt, B.P., Suntzeff, N.B., Phillips, M.M., Schommer, R.A., Clocchiatti, A., Kirshner, R.P., Garnavich, P., Challis, P., Leibundgut, B.R.U.N.O., Spyromilio, J. and Riess, A.G., 1998. The high-Z supernova search: measuring cosmic deceleration and global curvature of the universe using type Ia supernovae. *The Astrophysical Journal*, *507*(1), p.46.